\documentclass[pre,aps,twocolumn,superscriptaddress]{revtex4-1}
\pdfoutput=1 
\usepackage{amssymb,epsfig,amsmath,graphicx,textcomp,float,color,cases,epstopdf}
\usepackage{soul}

\usepackage[normalem]{ulem}
\usepackage{float}

\usepackage[titletoc]{appendix}
\usepackage{amsmath}

\usepackage[colorlinks=true, urlcolor=blue, anchorcolor=blue, citecolor=blue,filecolor=blue,linkcolor=blue,menucolor=blue]{hyperref}

\newcommand{\RomanNumeralCaps}[1]
    {\MakeUppercase{\romannumeral #1}}

\begin{document}
	\author{Tal Agranov}
	\email{tal.agranov@mail.huji.ac.il}
	\affiliation{Racah Institute of Physics, Hebrew University of
		Jerusalem, Jerusalem 91904, Israel}
		\author{P. L. Krapivsky}
		\email{pkrapivsky@gmail.com}
		\affiliation{Department of Physics, Boston University, Boston, Massachusetts 02215, USA}
	\author{Baruch Meerson}
	\email{meerson@mail.huji.ac.il}
	\affiliation{Racah Institute of Physics, Hebrew University of
		Jerusalem, Jerusalem 91904, Israel}

\title{Occupation time statistics of a gas of interacting diffusing particles}
		
\begin{abstract}
The time which a diffusing particle spends in a certain region of space is known as the occupation time, or the residence time. Recently the joint occupation time statistics of an ensemble of non-interacting particles was addressed using the single-particle statistics. Here we employ the Macroscopic Fluctuation Theory (MFT) to study the occupation time statistics of many \emph{interacting} particles.  We find that interactions can significantly change the statistics and, in some models, even cause a singularity of the large-deviation function describing these statistics. This singularity can be interpreted as a dynamical phase transition. We also point out to a close relation between the MFT description of the occupation-time statistics of non-interacting particles and the level 2 large deviation formalism which describes the occupation-time statistics of a single particle.
\end{abstract}
\maketitle

\section{Introduction}
\label{intro}
The amount of time a Brownian particle spends in some region of space, known as the occupation time, or a residence time, is a key quantity in the description of Brownian motion. This quantity was extensively studied
\cite{levi,Levy,kac,IM65,spi,noam1,noam2,noam3,satya,eli,BM:book,benyopty,greben,berez,many,beni,Ray,Knight,benyopty2,singlefile,tridib1,Hugo16,Hugo16b,
Hugo18,simtwo,oldsurv,reactions,sid}, and it has many applications in different fields. The occupation time statistics displays interesting and often surprising properties. One well-known example is the arcsine law \cite{levi} for the occupation of the half line. More recent results include the non-monotonicity of the occupation time as a function of the diffusion coefficient \cite{benyopty}, non-ergodicity \cite{satya,eli}, and non-analyticity of the large deviation function, characterizing the occupation time of a driven Brownian particle \cite{Hugo16b,Hugo18}. One important application of the occupation-time statistics deals with diffusion-controlled chemical reactions. Consider a receptor, whose activity is proportional to the time during which signaling molecules stay in its vicinity. Then the occupation time can be used to evaluate the reaction rate \cite{many,noam1,noam2,reactions,simtwo}. There can be \emph{many} signaling molecules and the activity is proportional both to time and to the number of molecules in the vicinity of the receptor \cite{many,simtwo}. The ensuing many-body problem has been addressed for \emph{non-interacting} Brownian molecules, where the calculations are based on the single-particle statistics \cite{many,simtwo}.

In many biologically relevant situations such as in crowded living cell, inter-particle interactions can be significant \cite{bress}, and the single-particle approach breaks down. In this work we continue a previous line of research \cite{MVK,surv,AMabsorption,narrow,AMreview} and address this many-body problem by employing the Macroscopic Fluctuation Theory (MFT) \cite{MFTreview}.  The MFT proves to be useful also for non-interacting particles. The formalism is versatile and can be applied to systems of different geometries and dimensions. For simplicity, we focus on one-dimensional systems. We show that interactions between diffusing particles strongly affect the occupation time statistics. They introduce a nontrivial dependence of the statistics on the number of particles. Remarkably, they lead, for some diffusive lattice gas models, to non-analyticity in the large deviation function which characterizes the occupation time statistics. Non-analyticities of this type are usually classified as dynamical phase transitions. Other examples of dynamical phase transitions in diffusive lattice gases  can be found in Refs. \cite{ber,main2,hurtado,main,phase}.

Here is a plan of the remainder of the paper. In Sec.~\ref{desc} we briefly discuss the single-particle occupation statistics and two existing formalisms for addressing them. We then define the occupation fraction of an \emph{ensemble} of diffusing particles and show how to calculate this quantity for non-interacting particles, using the single-particle statistics. In Sec.~\ref{fluc} we formulate the MFT of the occupation time statistics. In Sec. \ref{RWs} we test our theory for non-interacting random walkers (RWs) and reproduce the single-particle theory \cite{Hugo16b,Hugo18}. We also point out, for the non-interacting particles, to a close relation between the MFT of RWs and the level 2 single-particle large deviation formalism. In Secs.~\ref{ssep} and \ref{zrp}, we apply the MFT to two models of \emph{interacting} lattice gases:  the simple symmetric exclusion process (SSEP) and the zero-range process (ZRP). In Sec. \ref{ring}
we study the occupation statistics in finite systems and uncover a dynamical phase transition for a class of interacting lattice gases. In Appendix \ref{ringrws} we consider the role of finite-size effects in the occupation statistics of non-interacting particles. To our knowledge, previous studies of the finite-size effects dealt only with a single particle  and were limited to the evaluation of the variance of the occupation fraction \cite{berez}. Our main findings are summarized in Sec.~\ref{disc}.

\section{Occupation fraction: from a single particle to many}
\label{desc}
Let us consider a Brownian particle on the infinite line and denote by $X(t)$ its position at time $t$. The occupation fraction of the interval $|x|<l$ is the fraction of time $\nu$ that the particle spends inside this interval during the time interval $(0,T)$. This quantity is also known as the \emph{empirical measure} of the particle inside the interval:
\begin{equation}
\nu=\frac{1}{T}\int_0^Tdt\int_{-l}^{l}dx\,\delta\left[x-X(t)\right],
\label{const1}
\end{equation}
where $\delta\left(\dots\right)$ is the Dirac delta function. As the particle wanders along the line, it becomes exceedingly improbable for it to spend a finite fraction of time inside the interval, so that  the expected value $\bar{\nu}$ tends to $0$
as $T\to\infty$. The probability ${\mathcal P}_{1}(\nu,T\to \infty)$ of observing any finite value of $\nu$ has the large deviation form \cite{hugo2009}:
\begin{equation}
\mathcal P_{1}(\nu,T\to \infty)\sim e^{-Ts_{1}\left(\nu\right)}, \label{hugo}
\end{equation}
where the decay rate $s_{1}\left(\nu\right)$ plays the role of the large deviation function. As an extreme, $\mathcal P_{1}(\nu=1,T\to \infty)$ describes the probability that the Brownian particle remains inside the interval during the entire time $T$. This probability is known as the survival probability, and it was calculated long ago \cite{oldsurv,Paulbook}:
\begin{equation}
s_1\left(\nu=1\right)=\frac{\pi^2D_0}{4l^2},\label{PT}
\end{equation}
where $D_0$ is the diffusion coefficient of the Brownian particle. The calculation involves solving the diffusion equation with absorbing boundary conditions at $|x|=l$. In the long-time limit the solution is dominated by the smallest eigenvalue $\pi^2/4l^2$ of the Laplace operator with Dirichlet boundary conditions, which determines the decay rate (\ref{PT}).

The problem of computing the rate function $s_1(\nu)$ over its entire range $0\leq\nu\leq1$ is more involved, and it was addressed only recently \cite{Hugo16,Hugo16b,Hugo18,notehugo}. Within the Donsker-Varadhan (DV) large-deviation formalism \cite{DonskerVaradhan}, $s_1$ can be extracted from the cumulant generating function (the latter is the Legendre-Fenchel transform of the former). The generating function is obtained by solving a parameter-dependent eigenvalue problem for the so-called tilted generator \cite{Hugo16}. Here we will reproduce the single-particle solution of Refs. \cite{Hugo16,Hugo16b,Hugo18} and provide an analytic expression for the rate function (\ref{hugo}) in a parametric form. We will do it, however,  by specializing the MFT formalism to a system of many non-interacting particles. These
calculations will demonstrate a close connection between the MFT and two other methods: the DV formalism and
the level 2 large deviations formalism \cite{Hugo16,Hugo18,level2}.  The level 2 formalism was originally developed for equilibrium steady states, but it can be extended to non-stationary processes \cite{level2}.
The level 2 formalism deals with the rate functional $I[\rho_{1}(x)]$ which characterizes the distribution of the fluctuating empirical density of the particle:
\begin{equation}
\rho_{1}(x)=\frac{1}{T}\int_0^Tdt\,\delta[x-X(t)].
\end{equation}
The quantity $\rho_{1}(x)$ can be interpreted as a single-particle analog of a (time averaged)  number density of a ``gas" composed of only one particle. At long times the distribution of $\rho_{1}(x)$ obeys a large-deviation form
\begin{equation}
\mathcal P_{1}\left[\rho_{1}(x)\right]\sim e^{-TI[\rho_{1}(x)]}.
\label{hugolevel2}
\end{equation}
The decay rate $s_1(\nu)$, entering Eq.~(\ref{hugo}), is given by the minimum value of the rate functional $I$ when
minimized over all gas density profiles $\rho_1(x)$ which satisfy the occupation-fraction constraint
\begin{equation}
\int_{-l}^{l}dx\,\rho_{1}(x)=\nu
\label{const2}
\end{equation}
and the normalization constraint which ensures that the entire gas is composed of exactly one particle:
\begin{equation}
\int_{-\infty}^{\infty}dx\,\rho_{1}(x)=1.
\label{mass1}
\end{equation}
The optimal profile $\rho_{1}(x)$ plays an important role in the context of the conditioned process: the diffusion process, conditioned on realizing the specified occupation fraction $\nu$. In the long time limit, $\rho_{1}(x)$ is the position distribution of the particle conditioned on $\nu$ \cite{Hugo18,Hugo16,conditioned,conditioned2}. We will return to this property in Sec.~\ref{RWs}.

What happens when there are many diffusing particles on the line? It is natural to define the occupation fraction of $N$ identical particles as the one-particle occupation fraction (\ref{const1}), averaged over all particles:
\begin{equation}
\nu=\frac{1}{N}\sum_{i=1}^{N}\nu_i,\label{gasfrac}
\end{equation}
where $\nu_1,\nu_2,\dots\nu_{N}$ are the individual occupation fractions  of the particles. This is also the time averaged mass of particles inside the interval, normalized by their total mass. One reason to choose this definition is its relation to diffusion-limited reactions, as explained in the Introduction. Another reason has to do with experiment as we shall discuss in Sec.~\ref{disc}.

Let us first assume that the system is composed of \emph{non-interacting} Brownian particles. It is clear that the expected occupation fraction $\bar{\nu}$ converges to $0$ at long times $T$, meaning that the interval is almost empty during most of the time $T$. The extreme case $\nu=1$ corresponds to the so called survival problem, where \emph{all} the particles are conditioned to stay inside the interval for the entire time $T$. This case has already been considered for a system of many non-interacting and  interacting particles  \cite{surv,narrow}. A natural next question concerns the complete distribution of the occupation fraction, $0\leq\nu\leq1$. For the non-interacting particles, this quantity can be obtained from the single-particle distribution, see Eq.~(\ref{hugo}), recently derived in Ref.~\cite{Hugo16b,Hugo18}. The joint distribution of the set of individual occupation fractions $\{\nu_i\}$ is the product of the single-particle probabilities (\ref{hugo})
\begin{equation}
\mathcal P(\{\nu_i\})\sim e^{-T\sum_{i=1}^{N}s_{1}\left(\nu_i\right)}\label{hugo2}.
\end{equation}
The probability of an occupation fraction $\nu$ can be obtained by integrating (\ref{hugo2}) over the individual occupation fraction sets $\{\nu_i\}$ in the hyperplane defined by the constraint (\ref{gasfrac}). The integral can be evaluated using Laplace's method, and the resulting probability decays exponentially in time,
\begin{equation}
	\mathcal P(\nu,N)\sim e^{-Ts\left(\nu,N\right)},
\label{gas}
\end{equation}
with $s$ given by the minimum of the sum $\sum_{i=1}^{N}s_{1}\left(\nu_i\right)$ under the constraint (\ref{gasfrac}). Since $s_{1}\left(\nu_i\right)$ is a convex function \cite{Hugo16b,Hugo18}, this minimum is unique and it is given by equal individual occupation fractions $\nu_i=\nu$ for all $i$. Therefore
\begin{equation}
	s(\nu,N)=Ns_{1}(\nu)
\label{singlerw}
\end{equation}
showing that the probability of a specified occupation fraction (\ref{gasfrac}) of a system of non-interacting particles comes from equal contributions of all particles.

The main goal of this paper is to address the occupation statistics of \emph{interacting} particles. In this case the single-particle picture breaks down, and a different approach is required. Assuming that there are many particles in relevant regions of space, we employ a coarse-grained description given by \emph{fluctuating hydrodynamics} of diffusive lattice gases \cite{Spohn} and a corresponding large deviation theory known as the Macroscopic Fluctuation Theory (MFT) \cite{MFTreview}. We determine the distribution of the occupation fraction of the ensemble of particles and show that the exponential-in-time decay of the probability~(\ref{gas}) holds for a whole class of interacting particle systems. However, due to the interactions, the decay rate $s$ depends on the total particle number $N$ in a nontrivial way. Further, we identify a class of interacting particles for which the decay rate $s(\nu,N)$ is a non-analytic function of $\nu$. Such non-analyticities can be interpreted as dynamical phase transitions. Our formalism is also useful for non-interacting particles where it reproduces Eq.~(\ref{singlerw}) and recent results of Refs.~\cite{Hugo16b,Hugo18}.

\section{Macroscopic fluctuation theory of occupation statistics}
\label{fluc}

The starting point for the MFT (see the review \cite{MFTreview} for details) is the fluctuating hydrodynamics: a coarse-grained, in space and in time, description of a system composed of many diffusing particles in terms of the fluctuating particle number density $\rho(x,t)$, which obeys, in one dimension, the conservative Langevin equation \cite{Spohn,KL}:
\begin{equation}
\partial_t \rho = -\partial_x J ,\quad J=-D(\rho)\partial_x \rho-\sqrt{\sigma(\rho)}\eta\left(x,t\right),\label{lang}
\end{equation}	
where $D(\rho)$ and $\sigma(\rho)$ are the diffusivity and mobility of the ensemble of particles, respectively, and $\eta(x,t)$ is a zero-mean Gaussian noise, delta-correlated in space and time.

Equation \eqref{lang} is a stochastic partial differential equation. It provides a coarse-grained description for various transport models \cite{meso,jor,main,penini,shapiro2}. Here we will consider diffusive lattice gases \cite{Spohn,Liggett}: a family of models of particles hopping on a lattice. The simplest model of this type is the non-interacting random walkers (RWs) where each particle hops to neighboring sites with equal rates. For this model $D(\rho)=D_0$ and $\sigma(\rho)=2D_0\rho$. The coarse-grained behavior of the RWs coincides with that of non-interacting Brownian particles \cite{Paulbook}. Two \emph{interacting} lattice gases, which we will focus on, are the simple symmetric exclusion process (SSEP) and the Zero Range Process (ZRP). For the SSEP, only hops to empty neighboring sites are allowed,  and the transport coefficients are $D(\rho)=D_0$ and $\sigma(\rho)=2D_0\rho(1-a\rho)$ \cite{Spohn,KL}. We will set the lattice constant $a$ to unity, so that $0\leq\rho\leq1$. The SSEP mimics transport of hard particles in zeolites and biological channels \cite{Chou99},  water transport inside carbon nanotubes \cite{Das10}, various cellular processes \cite{bress,cell,oshanintracer}, and it has also been used to describe the transport of noninteracting electrons in mesoscopic materials at zero temperature \cite{jor}. The ZRP describes interacting particles without exclusion and with a vanishing interaction range, as here the hopping rate $w(n_i)$ depends only on the current occupation $n_i$ of the departure site. The ZRP also mimics different phenomena, such as shaken granular materials and growing networks, see \textit{e.g.} \cite{zrpbook} and references therein. For the ZRP the mobility is given by $\sigma(\rho)=2w\left(\rho\right)$, and the diffusivity is given by $D(\rho)=dw\left(\rho\right)/d\rho$ \cite{Spohn,void}. Formally, the RWs can be viewed as a particular case of the ZRP, where the hopping rate is proportional to the number of particles in the departure site.

We will introduce the MFT formalism for a large but finite number of particles
\begin{equation}
\int_{-\infty}^{\infty} dx\rho =N,\label{mass}
\end{equation}
on the whole line,  and later on modify it to account for systems of finite size. The Langevin equation (\ref{lang}) defines a path integral representation for the probability $\mathcal P$ of observing a joint density and flux histories $\rho(x,t), J(x,t)$, constrained by the conservation law (\ref{lang})
\begin{subequations}
\begin{align}
\label{path0}
&\mathcal P = \int\mathcal{D}\rho\mathcal{D}J
\prod_{x,t}\delta(\partial_{t}{\rho}+\partial_x J)\,
e^{-\mathcal{S}}\,,\\
&\mathcal{S}\left[\rho,J\right]=\int_0^Tdt\int_{-\infty}^{\infty} dx\frac{\left[J+D(\rho)\partial_x \rho\right]^2}{2\sigma(\rho)}\,.
\label{path1}
\end{align}
\end{subequations}
The path integration measure $\mathcal{D}\rho\mathcal{D}J$ accounts for the Jacobian of the transformation from $\eta$ to $\rho$ and $J$. We do not discuss its explicit form, as its account would only give a sub-leading contribution which is not captured by the MFT.

The probability ${\mathcal P}(\nu)$ of observing an occupation fraction $\nu$ is given by the path integral \eqref{path0} and \eqref{path1} evaluated over those histories which result in the specified value of $\nu\in[0,1]$. The latter can be expressed in terms of the particle number density $\rho(x,t)$:
\begin{equation}
\nu=\frac{1}{TN}\int_0^Tdt\int_{-l}^{l}dx\rho(x,t) .
\label{cons}
\end{equation}
Employing the number of particles in the relevant regions of space as
a large parameter  \cite{sad}, the MFT performs a saddle point evaluation of the path integral. The desired probability is mostly contributed to by the \emph{optimal} history of the system -- the most probable density and flux histories, leading to the specified value of $\nu$. The minimum action $S$, evaluated over these, yields the probability ${\mathcal P}\left(\nu,N\right)$ up to a pre-exponential factor:
\begin{eqnarray}
\label{actionmain}
-\ln {\mathcal P}
\simeq S \equiv\min_{\rho,J}\mathcal{ S}\left[\rho(x,t),J(x,t)\right].
\end{eqnarray}
If the averaging time $T$ is much longer than the characteristic diffusion time through the interval of length $2l$, the optimal profiles of $\rho(x,t)$ and $J(x,t)$ become stationary. In the context of fluctuations of current in lattice gases, driven by density reservoirs at the boundaries, this simple property is known as the additivity principle \cite{bd}. The additivity principle was verified in the survival limit $\nu =1$ \cite{surv}. Here we assume that it also holds for all $0<\nu<1$, and discuss the validity of this assumption in Sec. \ref{disc}.

In an infinite system with zero current at infinity, like our system, stationarity implies a zero stationary optimal current $J$. This means that the fluctuational contribution to the optimal flux exactly counterbalances the deterministic contribution and maintains a stationary density profile. Setting $J=0$ and $\rho=\rho(x)$ in Eq.~(\ref{path1}), we see that the action (\ref{actionmain}) is proportional to time, $S=Ts$, reproducing and generalizing Eq.~(\ref{gas}).  The action rate $s$ is obtained by minimizing the action rate functional
\begin{equation}\label{sq}
\mathfrak{s}\left[\rho\left(x\right)\right]=\int_{-\infty}^{\infty} dx\frac{\left[D(\rho)\partial_x \rho\right]^2}{2\sigma(\rho)}
\end{equation}
subject to the occupation fraction constraint
\begin{equation}
\int_{-l}^{l}dx\rho(x)=N\nu,\label{cons1}
\end{equation}
which follows from Eq.~(\ref{cons}). An additional constraint comes from mass conservation: the total number of particles on the line must be equal to the initial number $N$. This means that the total number of particles, composing the stationary density profile, cannot exceed the initial number of particles $N$. One could hypothesize that, at a given $\nu<1$,  some of the excess particles can diffuse away to infinity. As we will show, such a scenario is impossible for the lattice gas models considered here  \cite{Itwould}. That is, the optimal stationary density profile includes all $N$ particles, as described by Eq.~(\ref{mass}).

 Finally, the density profile $\rho$ must be everywhere nonnegative:
\begin{equation}
\rho\left(x\right)\geq 0.\label{positive}
\end{equation}
As we will see in Sec. \ref{ring}, this obvious constraint leads, for a class of interacting lattice gas models, to a dynamical phase transition.

Upon rescaling the spatial coordinate $x/l\rightarrow x$,  the occupation fraction constraint (\ref{cons1}) becomes
\begin{equation}
\int_{-1}^{1}dx\rho(x)=2n\nu,\label{cons2}
\end{equation}
where $n= N/2l$. The rescaled mass constraint (\ref{mass}) is
\begin{equation}
\int_{-\infty}^{\infty} dx\rho(x)=2n.\label{mass2}
\end{equation}
Then Eq.~(\ref{sq}) predicts the $1/l$ scaling of the action rate with the interval length $l$:
\begin{equation}\label{lag2}
s\left(\nu,N;l\right)=\frac{1}{l}\tilde{s}\left(\nu,n\right).
\end{equation}
The minimization problem for the functional~(\ref{sq}), subject to the integral constraints~(\ref{cons2}) and (\ref{mass2}), look simpler in the new variable  $u(x)=f\left[\rho\left(x\right)\right]$, where \cite{surv,MVK,Convergence}
\begin{equation}
f(\rho)=\int_0^\rho dz \frac{D(z)}{\sqrt{\sigma (z)}}.\label{transform1}
\end{equation}
As $D(z)$ and $\sigma(z)$ are non-negative, $f(\rho)$ monotonically increases with $\rho$. As a result, the inverse function $f^{-1}$, which we denote by $F(u)$, monotonically increases with $u$ in the relevant region of $u$.
In terms of $u(x)$, the action rate~(\ref{sq}) has the form of effective electrostatic energy
\begin{equation}
\label{su}
\mathfrak{s}\left[u\left(x\right)\right]=\frac{1}{2}\int_{-\infty}^{\infty} dx\left(\frac{du}{dx}\right)^2,
\end{equation}
which is universal for all interacting particle models, described by Eq.~(\ref{lang}). The constraints (\ref{cons2}) and (\ref{mass2}) are enforced via two Lagrange multipliers, which results in the Euler-Lagrange equation
\begin{numcases}
{\!\! u_{xx}=}
-\Lambda_1^2\frac{dF\left(u\right)}{du},   & $|x|\leq1$, \label{lagrangein}\\
\Lambda_2^2\frac{dF\left(u\right)}{du},   & $|x|\geq1$. \label{lagrangeout}
\end{numcases}
This choice of the signs of the interior and exterior Lagrangian multipliers turns out to be correct in all examples we will be dealing with.
We must also demand continuity of $u$ and $u_x$ at $x=1$. Finally, if there are multiple solutions, the one with the least action (\ref{su}) must be selected.

For some functions $F(u)$, Eq.~(\ref{lagrangeout}) can have a solution with compact support, $1<|x|<x_0$, where the first derivative $u'(x)$ vanishes at the edges of support, $|x|=x_0$, and the Lagrange multipliers $\Lambda_1^2$ and $\Lambda_2^2$ can be chosen to obey the constraints~(\ref{cons2}) and~(\ref{mass2}). In this case we simply set $u=0$ at $|x|>x_0$, which costs a zero action, see Eq.~(\ref{su}). The resulting optimal solution $u(x)$ obeys the tangent construction (the continuity of the first derivatives) at $|x|=x_0$. It can be considered as a solution of a one-sided variational problem \cite{tangent}, which results from the non-negativity constraint $u(x)\geq 0$, directly following from Eq.~(\ref{positive}). As we will see below, the presence of optimal solutions with compact support gives rise to a dynamical phase transition in finite systems. It is important, therefore, to find out whether or not, for a specified transport coefficients $D(\rho)$ and $\sigma(\rho)$, the solution has compact support and obeys the tangent construction. Equation~(\ref{lagrangeout}) has the ``energy" integral:
\begin{equation}
\frac{1}{2}\left( u_{x}\right)^2-\Lambda_2^2F\left(u\right) =
E ,\quad |x|>1.\label{energy}
\end{equation}
As $u$ and $u_x$ must vanish at infinity \cite{For}, so that the total mass (\ref{mass2}) is finite, the energy $E$ must be set to zero. Integrating Eq.~(\ref{energy}) over $x$ from $x=1$ up to $x_0$ where the solution vanishes, we arrive at
\begin{equation}
x_0=1+\frac{1}{\sqrt{2\Lambda_2^2}}\int_0^{u_1}\frac{dz}{\sqrt{F\left(z\right)}},\label{sup}
\end{equation}
where $u_1=u(x=1)$.
Compact support, $x_0<\infty$, demands that the integral in Eq.~(\ref{sup}) converge. The convergence is determined by the behavior of $F\left(u\right)$ in the vicinity of $u=0$ where $F\left(u\right)$ vanishes, see Eq.~(\ref{transform1}). Suppose that $D(\rho\rightarrow0)\sim\rho^{\alpha}$ and $\sigma(\rho\rightarrow0)\sim\rho^{\beta}$. Then for small $u$ one has $F(u)\sim u^{2/\left(2\alpha-\beta+2\right)}$, and the integral converges at $u\rightarrow0$ if and only if $2\alpha-\beta+1>0$. This condition [which is stricter than the condition $2\alpha-\beta+2>0$ that guarantees the existence of the function $F(u)$ at small $u$] is violated for the RWs and the SSEP, where $\alpha=0$ and $\beta=1$. For these two models, therefore, compact support of the optimal solution is impossible. For the ZRP, however, one has $D(\rho)\sim d\sigma(\rho)/d\rho$, so that $\alpha=\beta-1$. As a result, the integral  in Eq.~(\ref{sup}) converges for any hopping rate which grows with $\rho$ faster than linear ($\beta>1$, or equivalently $\alpha>0$). In this case the optimal profile has compact support. Note that, as $F(u=0)=0$ and $E=0$, the first derivative $u_x$ vanishes together with $u$ at $|x|=x_0$ [see Eq.~(\ref{energy})], so the tangent construction is satisfied.
Now we will consider the non-interacting RWs, where the support of the optimal solution is infinite.

\section{Non-interacting Random Walkers}\label{RWs}

For the RWs, the MFT minimization problem, defined by Eq.~(\ref{sq}), (\ref{cons2}) and (\ref{mass2}), \emph{coincides} with the minimization problem defined by the single-particle level 2 large deviation formalism. This is because the empirical density functional $I\left[\rho_{1}(x)\right]$ of the single-particle theory coincides with the action rate functional of the RWs, i.e. with $\mathfrak{s}\left[\rho(x)\right]$ given by Eq.~(\ref{sq}), see also \cite{level2}. Furthermore, the two constraints of the single-particle minimization, Eqs.~(\ref{const2}) and (\ref{mass1}), can be viewed as the gas constraints (\ref{cons2}) and (\ref{mass2}) applied to a gas composed of a single particle. This, together with the fact that, for the RWs, $s$ is proportional to $N$, immediately leads to Eq.~(\ref{singlerw}). Importantly, this also implies that the optimal particle number density $\rho\left(x\right)$, conditioned on an occupation fraction $\nu$, and the single particle's position distribution $\rho_{1}(x)$, conditioned on the same $\nu$, coincide up to factor $N$. This property generalizes a similar relation which holds at the level of the average behavior: The unconstrained average particle number density of the RWs coincides with the unconstrained position density distribution of a single particle up to the factor $N$.

The single-particle occupation statistics have been recently addressed in \cite{Hugo16b,Hugo18} within the DV formalism.  Before we turn to interacting particle systems, we will re-derive the results of Refs.~\cite{Hugo16b,Hugo18} by using the MFT formalism. For the RWs, Eq.~(\ref{transform1}) yields $F\left( u\right)=u^2/2D_0$, and Eqs.~(\ref{lagrangein}) and (\ref{lagrangeout}) turn into the Helmholtz equations
\begin{equation}
\label{helm}
u_{xx} =
\begin{cases}
-\lambda_1^2u       & |x|<1, \\
\lambda_2^2u        & |x|>1,
\end{cases}
\end{equation}
where $ \lambda^2=\Lambda^2/D_0$. Remarkably,  Eq.~(\ref{helm}) coincides,  up to relabeling of parameters,  with the eigenvalue problem for the tilted operator of the DV formalism for a single particle \cite{Hugo16b,Hugo18}. In particular, the field $u(x)$ coincides with each of the left and right eigenfunctions of the tilted operator. (The left and right eigenfunctions coincide in this case due to the hermiticity of the tilted operator.) Previously, equivalence was established between the tilted-generator formalism and the level 2 formalism \cite{Hugo18,conditioned2}. We thus conclude that for the RWs, the three formalisms are mathematically equivalent.	
As Eq.~(\ref{helm}) is  linear, the optimal profile (\ref{qrws}) scales linearly with $N$, and so does the action, as it should, see Eq.~(\ref{singlerw}).

\begin{figure}[h]
	\begin{tabular}{ll}
		\includegraphics[width=0.237\textwidth,clip=]{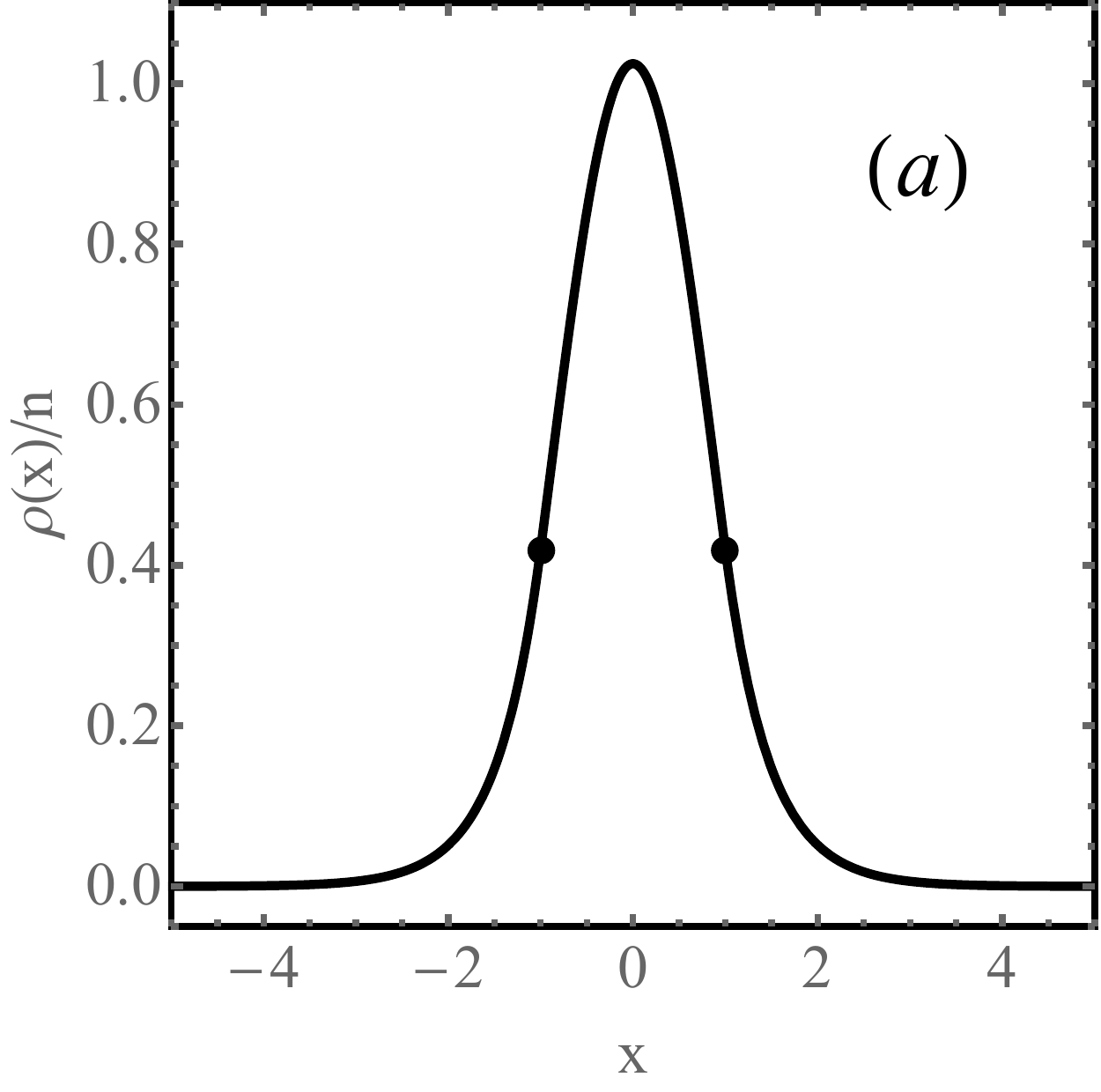}
		&
	\includegraphics[width=0.225\textwidth,clip=]{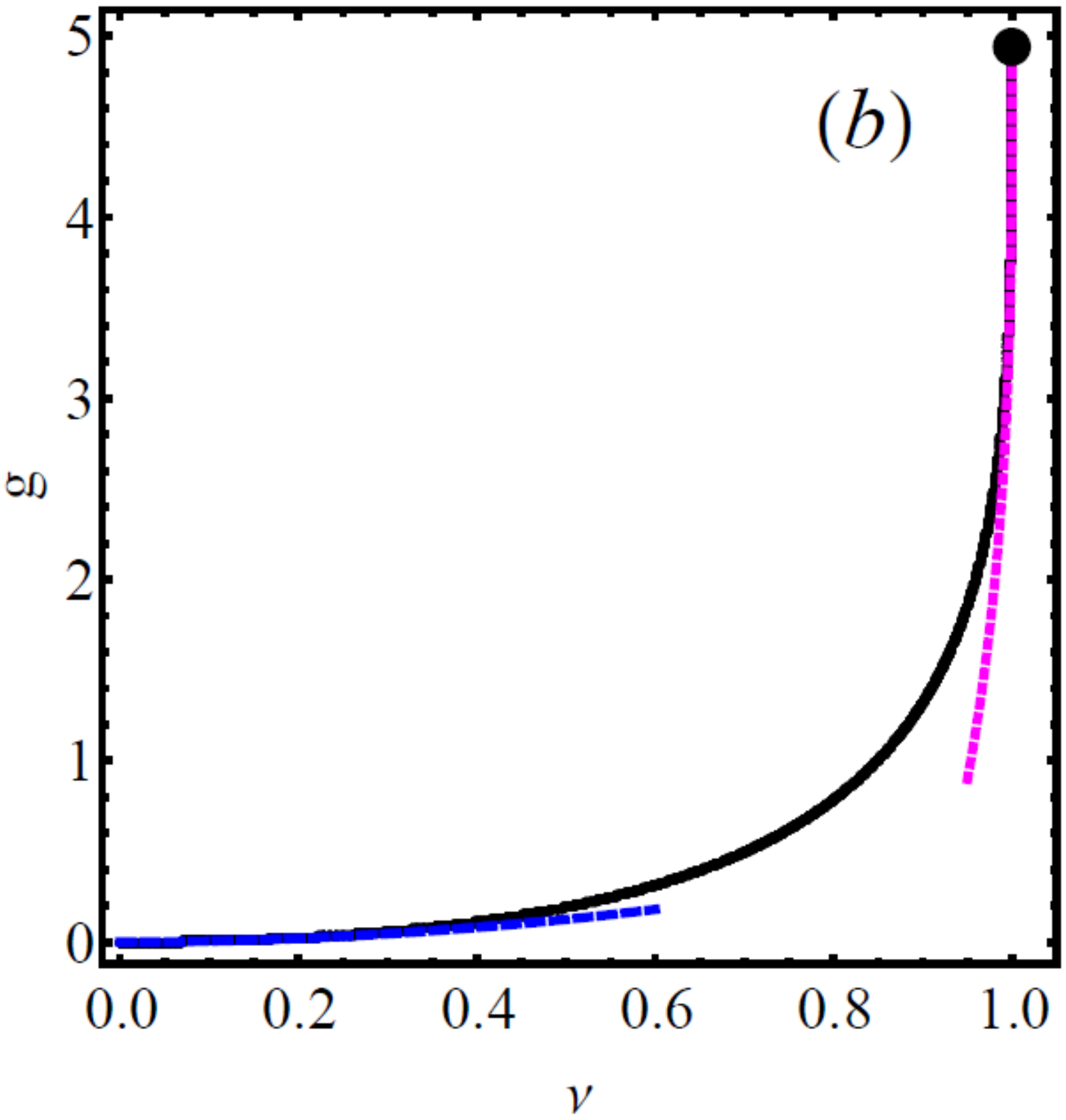}
	\end{tabular}	
	\caption{(a): The optimal profile $\rho(x)/n$ (\ref{qrws}) for independent RWs at $\nu=0.8$, which corresponds to $\lambda_1=0.875\dots$ as follows from Eq.~(\ref{para3}). The two fat dots mark the interval boundaries $|x|=1$. (b): The rescaled action $g(\nu)$, defined by (\ref{srws}) and (\ref{para3}), together with its asymptotics \eqref{low-high} at small and large occupation fraction (the dashed and dotted lines, respectively). The maximum value $g(\nu=1)=\pi^2/2$ corresponding to the survival limit~(\ref{PT}) is marked by the fat dot.}
	\label{srwsfig}		
\end{figure}

Up to relabeling of parameters, Eq.~(\ref{helm}) is the Schr\"{o}dinger equation for a finite square well potential, whose solution can be found in textbooks, see \textit{e.g.} Ref. \cite{quant}. The optimal (least-action) profile $u$ that we are after corresponds to the ground state of the quantum problem. This state is symmetric \cite{quant,landau}, and we obtain
\begin{equation}
\label{urws}
\frac{u}{\sqrt{2nD_0}} =
\begin{cases}
A\,\cos(\lambda_1 x)       & |x|<1, \\
B\,e^{-\lambda_2|x|}        & |x|>1,
\end{cases}
\end{equation}
where we have introduced a dimensional factor $\sqrt{2nD_0}$, so that the constants $A$ and $B$ are dimensionless.
What is left is to determine $A$, $B$ and the Lagrange multipliers $\lambda_1$ and $\lambda_2$. The continuity of the solution and its derivative at $|x|=1$ yields
\begin{equation}
\label{conti}
A\cos\left(\lambda_1\right)=Be^{-\lambda_2}\,, \quad A\lambda_1\sin\left(\lambda_1\right)=B\lambda_2e^{-\lambda_2}\,.
\end{equation}
The constraints (\ref{cons2}) and (\ref{mass2}) provide two more relations:
\begin{equation}
\label{rwsconst}
A^2\left[\frac{\sin\left(2\lambda_1\right)}{2\lambda_1}+1\right] =2\nu,\quad
\left(Be^{-\lambda_2}\right)^2 =2\lambda_2(1-\nu).
\end{equation}
The four algebraic equations~(\ref{conti}) and (\ref{rwsconst}) can be solved in a parametric form. It is convenient to use a single parameter $0\leq\lambda_1\leq\pi/2$ which corresponds to the whole occupation fraction range, $0\leq\nu\leq 1$ \cite{Actually}. The optimal profile $\rho(x)=u^2(x)/2D_0$ for a given value $\nu$ can be expressed in terms of the parameter $\lambda_1$:
\begin{equation}
\label{qrws}
\frac{\rho(x)}{n} =
\begin{cases}
A^2(\lambda_1)\,\cos^2(\lambda_1 x)                             & |x|<1, \\
B^2(\lambda_1)\,e^{-2|x| \lambda_1\tan\lambda_1}        & |x|>1,
\end{cases}
\end{equation}
where
\begin{subequations}
\begin{align}
\label{para1}
&A^2(\lambda_1) =  \frac{2\lambda_1}{\cot\lambda_1+\lambda_1}\,,\\
&B^2(\lambda_1) =
\frac{2\lambda_1\cos^2\lambda_1\,e^{2\lambda_1\tan\lambda_1}}{\cot\lambda_1+\lambda_1}\,,
\label{para2}
\end{align}
\end{subequations}
with $\lambda_1\left(\nu\right)$  implicitly determined by Eq.~(\ref{para3}) below. An example of the optimal profile is shown in Fig.~\ref{srwsfig}a.
Plugging Eq.~(\ref{urws}) into Eq.~(\ref{su}) [or Eq.~(\ref{qrws}) into Eq.~(\ref{sq})], and using Eq.~(\ref{actionmain}), we obtain
\begin{eqnarray}
-\ln\mathcal P(\nu,N)\simeq\frac{D_0TN}{2l^2}g(\nu)\label{actionrws}
\end{eqnarray}
where the function $g(\nu)$ is given in a parametric form:
\begin{eqnarray}
g(\lambda_1)&=&\frac{2\lambda_1^3}{\cot\lambda_1+\lambda_1},\label{srws}\\
\nu(\lambda_1)&=&\frac{\sin\lambda_1\cos\lambda_1+\lambda_1}{\cot\lambda_1+\lambda_1}.\label{para3}
\end{eqnarray}
The graph of function $g(\nu)$, shown in Fig.~\ref{srwsfig}b, agrees with the graph presented, without an analytic formula, in Refs. \cite{Hugo16b,Hugo18}. The maximum value of the action (the minimum probability) is obtained in the survival case $\nu=1$, or $\lambda_1=\pi/2$. The maximum value $g(\nu=1)=\pi^2/2$ is in agreement with Eq.~(\ref{PT}). By expanding Eq.~(\ref{para3}) near $\lambda_1=\pi/2$, one can obtain the asymptotic near the survival limit $\nu=1$.  The opposite asymptotic of small $\nu$ can be  obtained by expanding around $\lambda_1=0$. These lead to the explicit expressions
\begin{equation}
\label{low-high}
g(\nu)\simeq
\begin{cases}
\frac{\nu^2 }{2}         & \nu\ll 1, \\
\frac{\pi^2}{2}-\frac{3\pi^{4/3}}{2^{1/3}}\left(1-\nu\right)^{1/3}    &0<1-\nu\ll 1.
\end{cases}
\end{equation}
The latter asymptotic agrees with the corresponding result of Ref.~\cite{Hugo18}.

Now we are in a position to justify our choice of mass normalization~(\ref{mass}). The final total number of particles, conditioned on a specified $\nu$, is set by the minimum of the action (\ref{actionrws}) with respect to the total number of particles $N$. This corresponds to minimizing $N g\left(m/N\right)$ with respect to $N$ at fixed $m$. By virtue of the convexity of $g$, the minimum is achieved for the maximal possible total number of particles, which is equal to $N$.

\section{Simple Symmetric Exclusion Process }\label{ssep}

\subsection{General}
\label{SSEPgeneral}

\begin{figure}
	\includegraphics[width=0.3\textwidth,clip=]{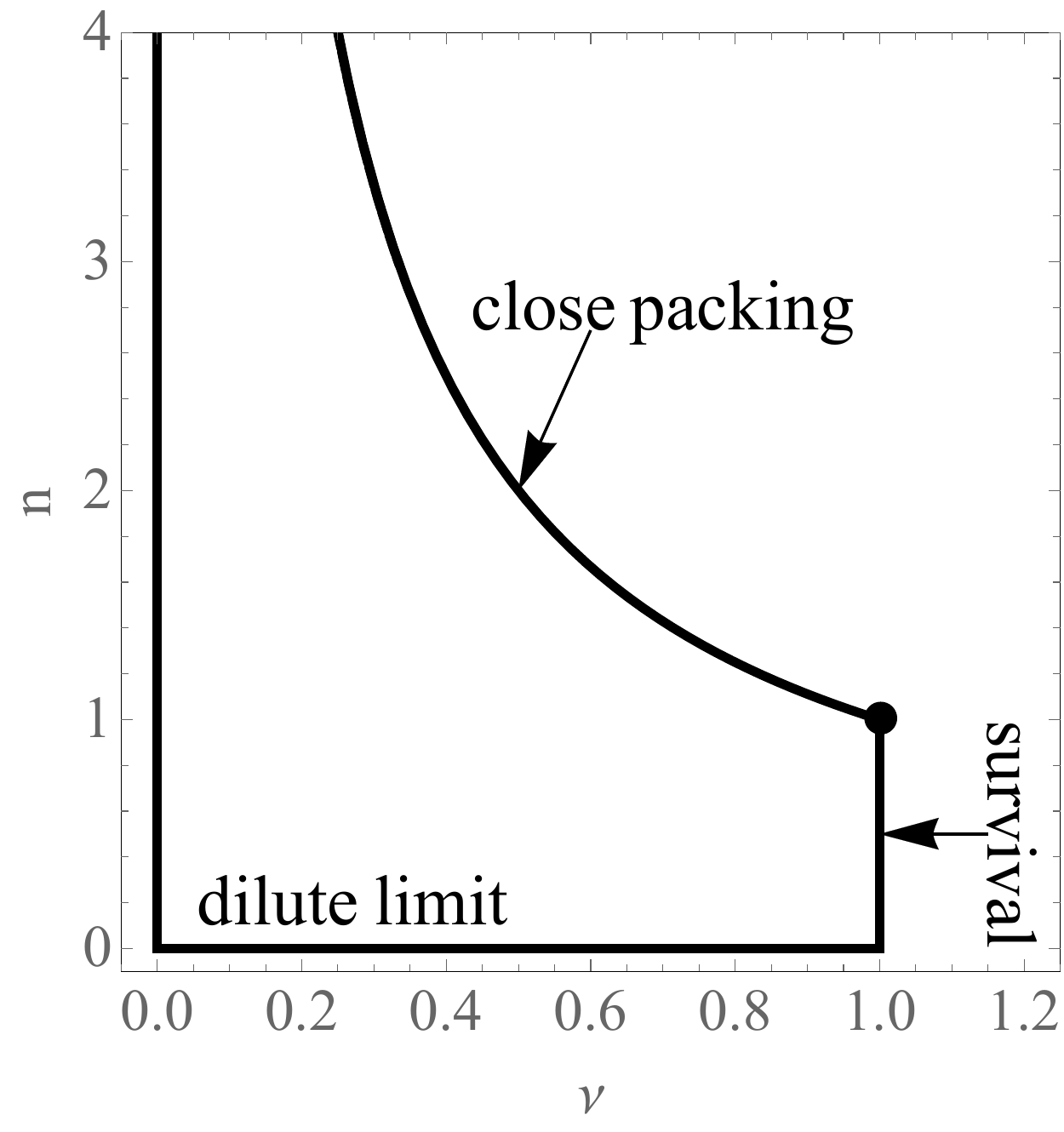}
	\caption{The parameter plane of the occupation statistics of the SSEP is bound by the close-packing hyperbola $\nu n=1$, the particle survival line $\nu=1$ and the axes $\nu$ and $n$. The circle marks the particle survival at the close packing point $(\nu=1,n=1)$.}
	\label{phase}
\end{figure}

The inter-particle interactions in the SSEP introduce a nontrivial dependence of the occupation statistics on the total number of particles. To begin with, the particle exclusion defines the boundaries of the parameter plane ($\nu, \, n$), see Fig.~\ref{phase}. The number of particles, that can occupy the interval $|x|<1$, is bounded from above, as the particle density cannot exceed the close packing value $\rho=1$. This yields the inequality $\nu n \leq 1$. We can still consider an arbitrary total number of particles in the system, assuming that not all of them were necessarily released inside the interval $|x|<1$. Finally, in the low-density limit $\nu n\rightarrow 0$ the inter-particle interactions are negligible, and we should expect to reproduce our results for  the RWs, Eqs.~(\ref{qrws}) and (\ref{para3}).

\begin{figure}[h]
	\begin{tabular}{ll}
		\includegraphics[width=0.23\textwidth,clip=]{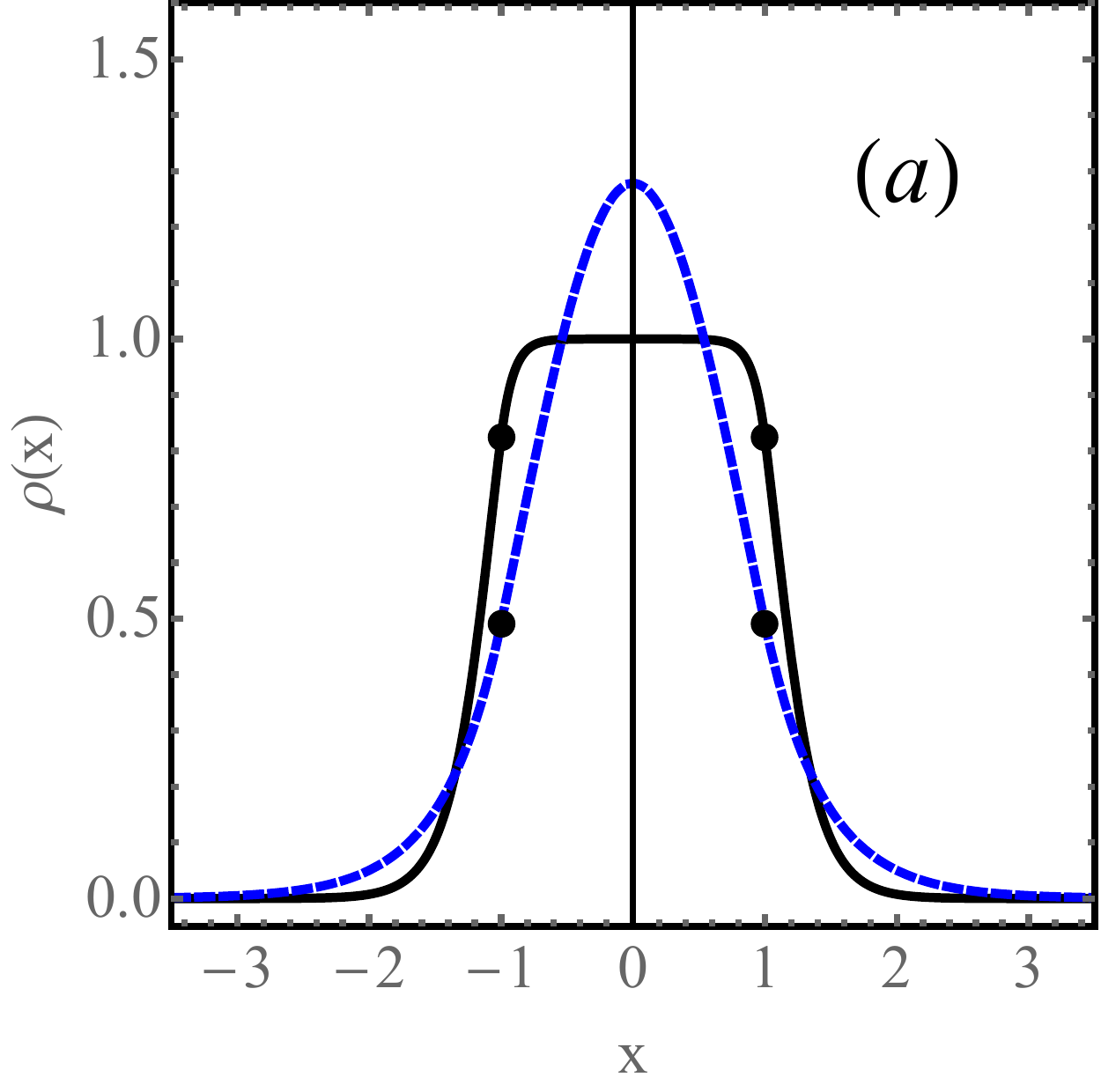}
		&
			\includegraphics[width=0.22\textwidth,clip=]{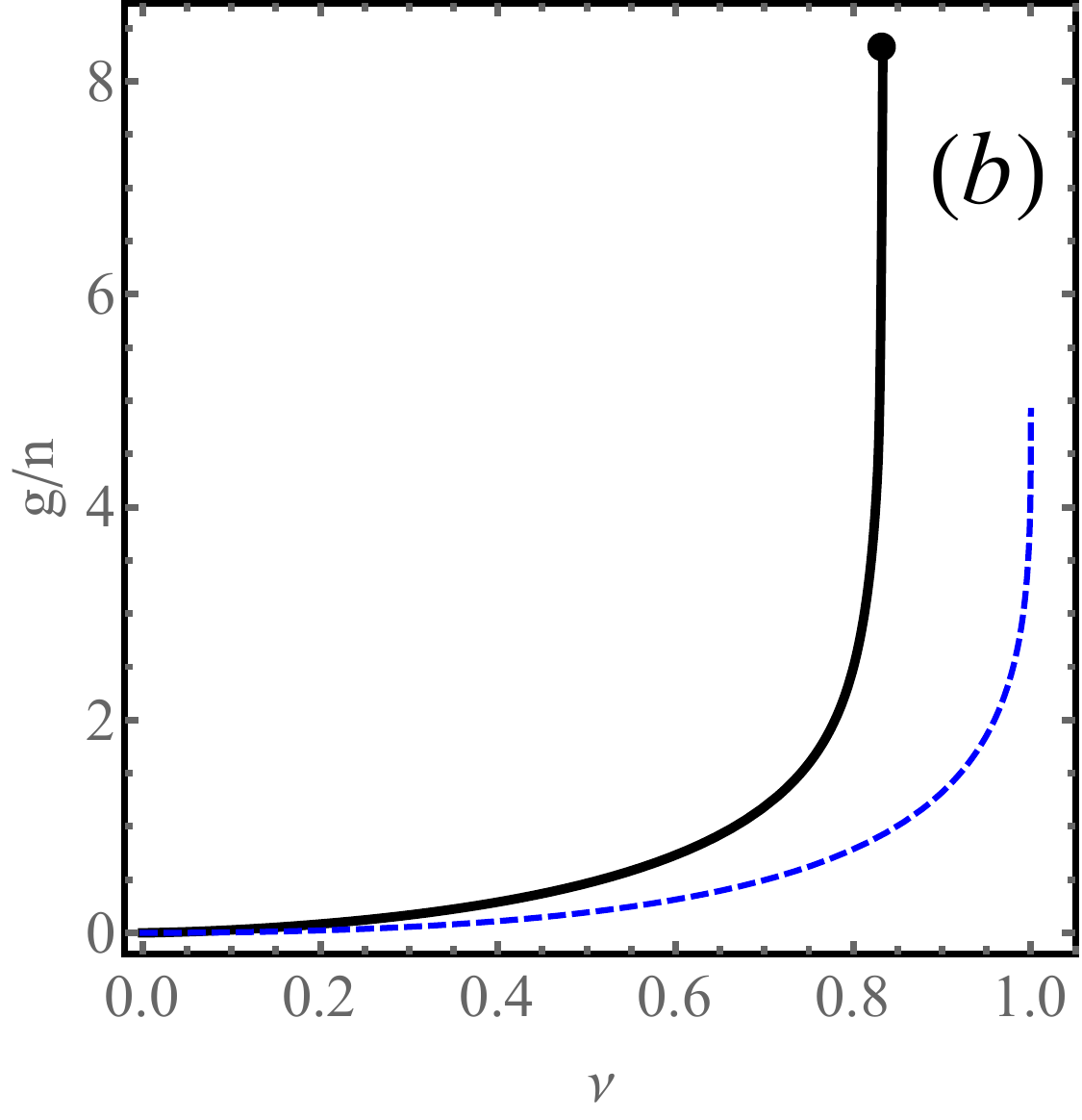}
	\end{tabular}	
	\caption{(a): The stationary optimal  profile $\rho(x)$ from Eq.~(\ref{qssep}) for $\nu=0.82$ and $n=1.2$ (solid line). For comparison, also shown in  dashed line is the profile (\ref{qrws}) for the RWs with the same parameter values. The circles mark the boundaries between the interior $|x|<1$ and the exterior $|x|>1$ parts of these solutions. For these parameters the maximum density of the SSEP is close (but still lower) than the close-packing value $\rho=1$. (b): The function $g(\nu,n)/n$, see Eq.~(\ref{sssep}), computed along the curve $n=1.2$ (solid line). The fat dot marks the close packing value (\ref{close}) $g(\nu=1/n,n)/n=8.33\dots$. For comparison, the dashed line depicts the RWs result $g(\nu)$, see Eq.~(\ref{srws}). }
	\label{rho}		
\end{figure}

For the SSEP, Eq.~(\ref{transform1}) yields $F(u)=\sin^2\left(u/\sqrt{2D_0}\right)$. Then, upon rescaling $U= \sqrt{2/D_0}\, u$ and $\lambda^2= \Lambda^2/D_0$, Eqs.~(\ref{lagrangein}) and~(\ref{lagrangeout}) turn into the stationary sine-Gordon equations \cite{surv}
\begin{equation}
\label{s12}
U_{xx}=
\begin{cases}
-\lambda_1^2\sin U   & |x|<1, \\
\lambda_2^2\sin U    & |x|>1.
\end{cases}
\end{equation}
Once $U(x)$ is determined, one can find
\begin{equation}
\rho(x)=F[u(x)]=\sin^2\left[\frac{U(x)}{2}\right].
\label{transssep1}
\end{equation}
Equations (\ref{s12}) are nonlinear, and the convenient Schr\"{o}dinger analogy is lost. Still, it is not difficult to solve them. The symmetric solution of (\ref{s12}), vanishing as $|x|\rightarrow\infty$ reads
\begin{equation}
\label{ussep}
U=
\begin{cases}
2\arcsin\left\{k\,\text{sn}\left[\lambda_1x+\text{K}(k),k\right]\right\}   & |x|<1, \\
4\arctan \left(e^{-\lambda_2|x|+A}\right)   & |x|>1,
\end{cases}
\end{equation}
where $\text{sn}(\dots)$ is the Jacobi elliptic function \cite{Wellipticfunctions}, and $\text{K}(\dots)$ is the complete elliptic integral of the first kind \cite{Wellipticintegrals}. As for the RWs, we have to determine the integration constants, $k$ and $A$ and the Lagrange multipliers $\lambda_1$ and $\lambda_2$. Imposing the continuity of $U'(x)$ at $|x|=1$, we can express $\lambda_2$ through $\lambda_1$ and $k$:
\begin{equation}
\lambda_2=\lambda_1\,\sqrt{1-k^2}\,\,\frac{\text{sn}\left(\lambda_1,k\right)}{\text{cn}\left(\lambda_1,k\right)}.\label{l2}
\end{equation}
Using this relation and the continuity $U(x)$ at $|x|=1$ we express $A$ in terms of $\lambda_1$ and $k$:
\begin{eqnarray}
A &=&\lambda_1\,\sqrt{1-k^2}\,\,\frac{\text{sn}\left(\lambda_1,k\right)}{\text{cn}\left(\lambda_1,k\right)} \nonumber \\
   &- &\ln\frac{1+\text{dn}\left[\lambda_1+\text{K}(k),k\right]}{k\,\text{sn}\left[\lambda_1+\text{K}(k),k\right]}\,,
\label{a2}
\end{eqnarray}
where $\text{cn}(\dots)$ and $\text{dn}(\dots)$ in Eqs.(\ref{l2}) and (\ref{a2}) are Jacobi elliptic functions \cite{Wellipticfunctions}.
The constants $\lambda_1$ and $k$ can be determined by using Eqs.~(\ref{cons2}) and~(\ref{mass2}). It is convenient to define two auxiliary expressions which involve $\lambda_1$ and $k$:
\begin{subequations}
\begin{align}
\label{q1}
&q_1=1-\frac{E\left[\text{am}\left(\lambda_1,k\right),k\right]-\frac{k^2\,
\text{cn}\left(\lambda_1,k\right)\,\text{sn}\left(\lambda_1,k\right)}
{\text{dn}\left(\lambda_1,k\right)}}{\lambda_1}\,,\\
\label{q2}
&q_2=\frac{1-\text{dn}\left[\lambda_1+\text{K}(k),k\right]}{\lambda_1\,\sqrt{1-k^2}}\,\frac{
\text{cn}\left(\lambda_1,k\right)}{\text{sn}\left(\lambda_1,k\right)}\,,
\end{align}
\end{subequations}
where $E(\dots,\dots)$ is the incomplete elliptic integral of the second kind and $\text{am}(\dots,\dots)$ is the Jacoby amplitude \cite{Wellipticfunctions,Wellipticintegrals}.
In terms of $q_1$ and $q_2$, the constraints (\ref{cons2}) and (\ref{mass2}) read:
\begin{equation}
n=q_1+q_2 \quad \text{and} \quad \nu=\frac{q_1}{q_1+q_2}.\label{qrho}
\end{equation}
As for the RWs, one cannot express $\lambda_1$ and $k$ through $\nu$ and $n$ in an explicit form. Moreover, even a parametric solution here demands two, rather than one, parameters. These two parameters, $0\leq k\leq1$ and $0\leq\lambda_1\leq\text{K}(k)$, correspond to the whole range $0\leq\nu\leq 1$ and $n>0$. The optimal profile (\ref{transssep1}) is given by
\begin{equation}
\label{qssep}
\rho(x)=
\begin{cases}
k^2\,\text{sn} ^2\left[\lambda_1x+\text{K}(k),k\right] & |x|<1, \\
\cosh^{-2}\left(\lambda_2|x|-A\right)  & |x|>1,
\end{cases}
\end{equation}
together with Eq.~(\ref{l2}) and (\ref{a2}), where the relations $k(\nu,n)$ and $\lambda_1(\nu,n)$ are given implicitly by Eqs.~(\ref{qrho}). An example of the optimal profile for $\nu=0.9$ and $n=1.1$ is shown in Fig.~\ref{rho}a.

Inserting Eq.~(\ref{ussep})  into Eq.~(\ref{su}) and using Eq.~(\ref{actionmain}), we obtain after some algebra \cite{Reintroducing}:
\begin{eqnarray}
-\ln\mathcal P(\nu,N,T)\simeq\frac{D_0T}{l}g\left(n,\nu\right),\label{actionssep}
\end{eqnarray}
where the function $g$ is given in a double parametric form:
\begin{eqnarray}
g\left(n,\nu\right)=2\left[\lambda_1^2\left(k^2-q_1\right)+\lambda_2^2q_2\right],\label{sssep}
\end{eqnarray}
together with Eqs.~(\ref{l2}), (\ref{q1}) and~(\ref{q2}) [where, again, the functions $k(\nu,n)$ and $\lambda_1(\nu,n)$ are given implicitly by Eq.~(\ref{qrho})].
As an example, Fig.~\ref{rho} shows the rescaled action $g(n,\nu))/n$  versus $\nu$ at $n=1.2$. For any $\nu$ and $n$, the action (\ref{actionssep}) for the SSEP is larger than that for the RWs (\ref{actionrws}). This is to be expected in view of the mutual exclusion of the SSEP particles. As one can check [by examining the action (\ref{actionssep}) in the same way as it was done in the end of Sec. \ref{RWs} for the RW], the normalization condition (\ref{mass}) holds for the SSEP.

Now we consider three different asymptotic behaviors of the action~(\ref{sssep}).

\subsection{Survival, $\nu=1$}

The survival limit $\nu=1$ was previously solved in Ref. \cite{surv}. Here $\rho(x)$ must vanish at $|x|\geq 1$. This demand sets $\lambda_1=\text{K}\left(k\right)$ \cite{surv}. Plugging this $\lambda_1$ into Eqs.~(\ref{l2}) and (\ref{a2}), we find that $\lambda_2\rightarrow\infty$ and $A\rightarrow\infty$ and obtain
\begin{equation}
\label{u1surv}
\rho(x)=
\begin{cases}
k^2\,\text{sn} ^2\left[\text{K}(k)\left(x+1\right),k\right]   & |x|<1, \\
0                                                                                   & |x|>1.
\end{cases}
\end{equation}
Plugging $\lambda_1=\text{K}\left(k\right)$ into Eqs.~(\ref{qrho}) we see that $\nu=1$ and
\begin{eqnarray}
n=1-\frac{\text{E}\left(k\right)}{\text{K}\left(k\right)},\label{parasurv}
\end{eqnarray}
where $\text{E}(\dots)$ is the complete elliptic integral of the second kind \cite{Wellipticintegrals}.
Using Eq.~(\ref{parasurv}) in Eq.~(\ref{sssep}), we obtain the action parametrized by a single parameter:
\begin{equation}
g\left(\nu=1,n\right)= 2\text{K}^2\left(k\right)\left[\frac{\text{E}\left(k\right)}{\text{K}\left(k\right)}+k^2-1\right],\label{actionsurv}
\end{equation}
where the function $k\left(n\right)$ is given by Eq.~(\ref{parasurv}). The action diverges at the special point $(\nu=1, n=1)$ of the parameter plane corresponds to survival at close packing, see Fig.~\ref{phase}. When approaching this point along the survival line $\nu=1$,   the asymptotic of the action can be obtained from Eq.~(\ref{actionsurv}) by expanding (\ref{parasurv}) near $k=1$. This  leads to
\begin{equation}
g\left(\nu=1,n\rightarrow1\right)\simeq \frac{2}{1-n}.\label{survpack}
\end{equation}
The results of this subsection agree with results of \cite{surv}.

\subsection{Close packing, $\nu n= 1$}
Here the interval $|x|<1$ is occupied to its maximum capacity, and we have $n\geq1$.  Survival at close packing point, where the MFT action diverges, can be reached along the close packing hyperbola $\nu n=1$ by taking the limit $n\rightarrow 1$. When $n>1$, that is whenever there are additional particles outside the interval, the action is finite. This regime is described by the limit of $k\rightarrow1$ of the general expressions of subsection \ref{SSEPgeneral}. It is more convenient, however, to directly solve Eq.~(\ref{s12}) for $|x|>1$ and match the solution to the close-packed solution $\rho=1$ for $|x|<1$. Then, using the mass constraint (\ref{mass2}), we obtain a simple expression for the optimal density profile:
\begin{equation}
\label{qssepclose}
\rho(x)=
\begin{cases}
1          & |x|<1, \\
\cosh^{-2} \left[\frac{2 \left(|x|-1\right)}{n-1}\right]      & |x|>1.
\end{cases}
\end{equation}
Plugging it  in Eq.~(\ref{sq}) and using the relation $S=s T$, we arrive at Eq.~(\ref{actionssep}) with the rescaled action
\begin{equation}
g\left(\nu=\frac{1}{n},n\right)= \frac{2}{n-1}.
\label{close}
\end{equation}
This action vanishes in the limit $n\rightarrow\infty$, as here the entire infinite line is closely packed, and so is the interval, with probability $1$.

\subsection{Dilute limit, $\nu n\ll1$}

In the dilute limit, $\nu n\ll1$, the exclusion effects of the SSEP are negligible, and one expects to arrive at
Eqs.~(\ref{qrws})-(\ref{para3}) for the independent RWs. This is indeed what happens (see Appendix \ref{rwsapend}) when we take the  limit  $k\rightarrow0$ in  Eqs.~(\ref{qssep}) and (\ref{sssep}). In particular, the rescaled action $g$ becomes
\begin{equation}
g_{\text{SSEP}}\left(\nu,n\right)\simeq n g_{\text{RWs}}\left(\nu\right).\label{rs}
\end{equation}

\section{Zero Range Process}\label{zrp}

As we showed in Sec. \ref{fluc}, for the ZRP model the stationary optimal density profile, conditioned on a given occupation fraction, has compact support when the hopping rate grows faster than linearly with $\rho$. For a power-law hopping rate the transport coefficients are
\begin{equation}
\label{ZRPcoeff}
D(\rho)=\Gamma\rho^{\alpha} ,\quad \sigma(\rho)=\frac{2\Gamma\rho^{\alpha+1}}{\alpha+1},
\end{equation}
where $\Gamma=\text{const}$. Here the only dimensional parameters of the problem are $T$, $n$, $l$ and $D(n)$. As we showed in Sec.\ref{fluc}, the action must be proportional to $T$ and inversely proportional to $l$ (\ref{lag2}). A simple dimensional analysis immediately yields the scaling
\begin{eqnarray}
	-\ln\mathcal P(\nu,N)\simeq\frac{D\left(n\right)n T}{l} g(\nu)=\frac{\Gamma N^{\alpha+1} T}{2^{\alpha+1}l^{\alpha+2}} g(\nu) ,\label{actzrp0}
\end{eqnarray}
and we only need to determine the rescaled action $g(\nu)$.  (Notice that
$\alpha=0$ corresponds to the RWs, where the action is proportional to $N$, and the density profile does not have compact support.) Dimensional analysis also implies that the optimal profile must be proportional to the density $n$.

Let us focus on the case $\alpha=1$, where the algebra is especially simple. Here the change of variables~(\ref{transform1}) is an identity, so we might as well solve for $\rho(x)$. Furthermore,  Eqs.~(\ref{lagrangein}) and~(\ref{lagrangeout}) simplify to
\begin{equation}
\label{3z}
\rho_{xx}=
\begin{cases}
-2\Lambda_1^2     & |x|<1, \\
2\Lambda_2^2       & 1 < |x|<x_0,
\end{cases}
\end{equation}
where $\pm x_0$  describe the edges of the compact support, see Sec.~\ref{fluc}. Equations~(\ref{3z}) are strikingly simple, and their symmetric solution is
\begin{equation}
\label{123}
\frac{\rho(x)}{n}=
\begin{cases}
\rho_{\text{max}}-\lambda_1^2x^2                & |x|<1, \\
\lambda_2^2\left(|x|-x_0\right)^2                  & 1<|x|<x_0, \\
0                                                                   & |x|>x_0,
\end{cases}
\end{equation}
where $\lambda=\Lambda/n$. The continuity of $\rho$ and $\rho_x$ at $|x|=1$ leads to
\begin{eqnarray}
\rho_{\text{max}}-\lambda_1^2&=&\lambda_2^2\left(1-x_0\right)^2,\label{100}\\
-\lambda_1^2&=&\lambda_2^2\left(1-x_0\right),
\end{eqnarray}
whereas Eqs.~(\ref{cons2}) and~(\ref{mass2}) yield
\begin{eqnarray}
2\left(\rho_{\text{max}}-\frac{\lambda_1^2}{3}\right)&=&\nu,\\
\frac{2}{3}\lambda_2^2\left(x_0-1\right)^3&=&1-\nu.\label{200}
\end{eqnarray}
Solving Eqs.~(\ref{100})--(\ref{200}) we obtain
\begin{equation}
\begin{split}
\lambda_1^2\left(\nu\right)&=\frac{3}{8}\left[9-5\nu-3\sqrt{\left(9-\nu\right)\left(1-\nu\right)}\right], \\
\lambda_2^2\left(\nu\right)&=\frac{3}{16}\left[\left(9-7\nu\right)\sqrt{\frac{9-\nu}{1-\nu}}+9\nu-27\right],\\
\rho_{\text{max}}\left(\nu\right)&=\frac{3}{8}\left(3+\nu-\sqrt{\left(9-\nu\right)\left(1-\nu\right)}\right),\\
x_0\left(\nu\right)&=\frac{3-\nu+\sqrt{\left(9-\nu\right)\left(1-\nu\right)}}{2\nu}.
\end{split}
\label{zrpparameter}
\end{equation}

\begin{figure}[h]
	\begin{tabular}{ll}
		\includegraphics[width=0.236\textwidth,clip=]{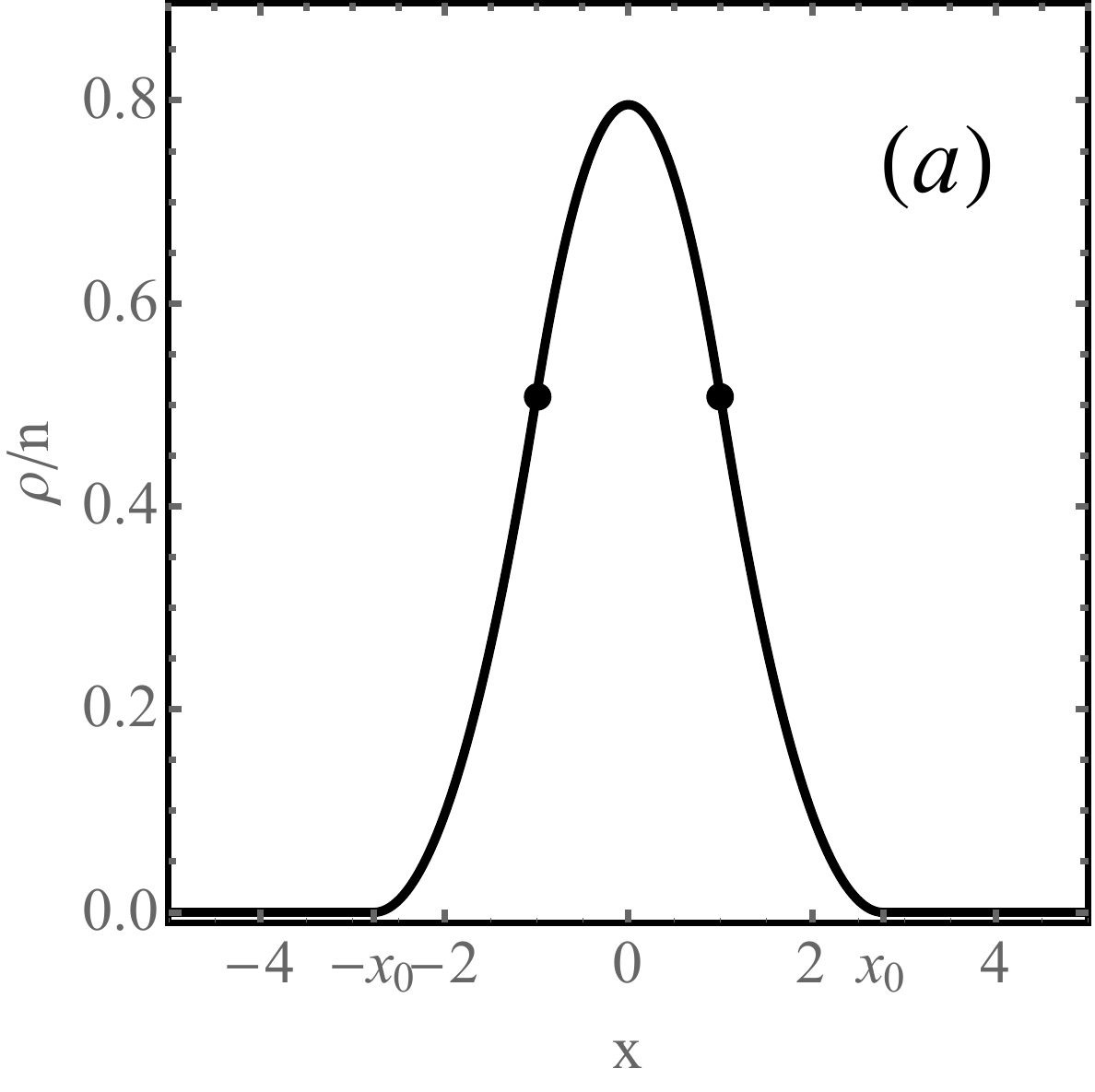}
		&
	\includegraphics[width=0.23\textwidth,clip=]{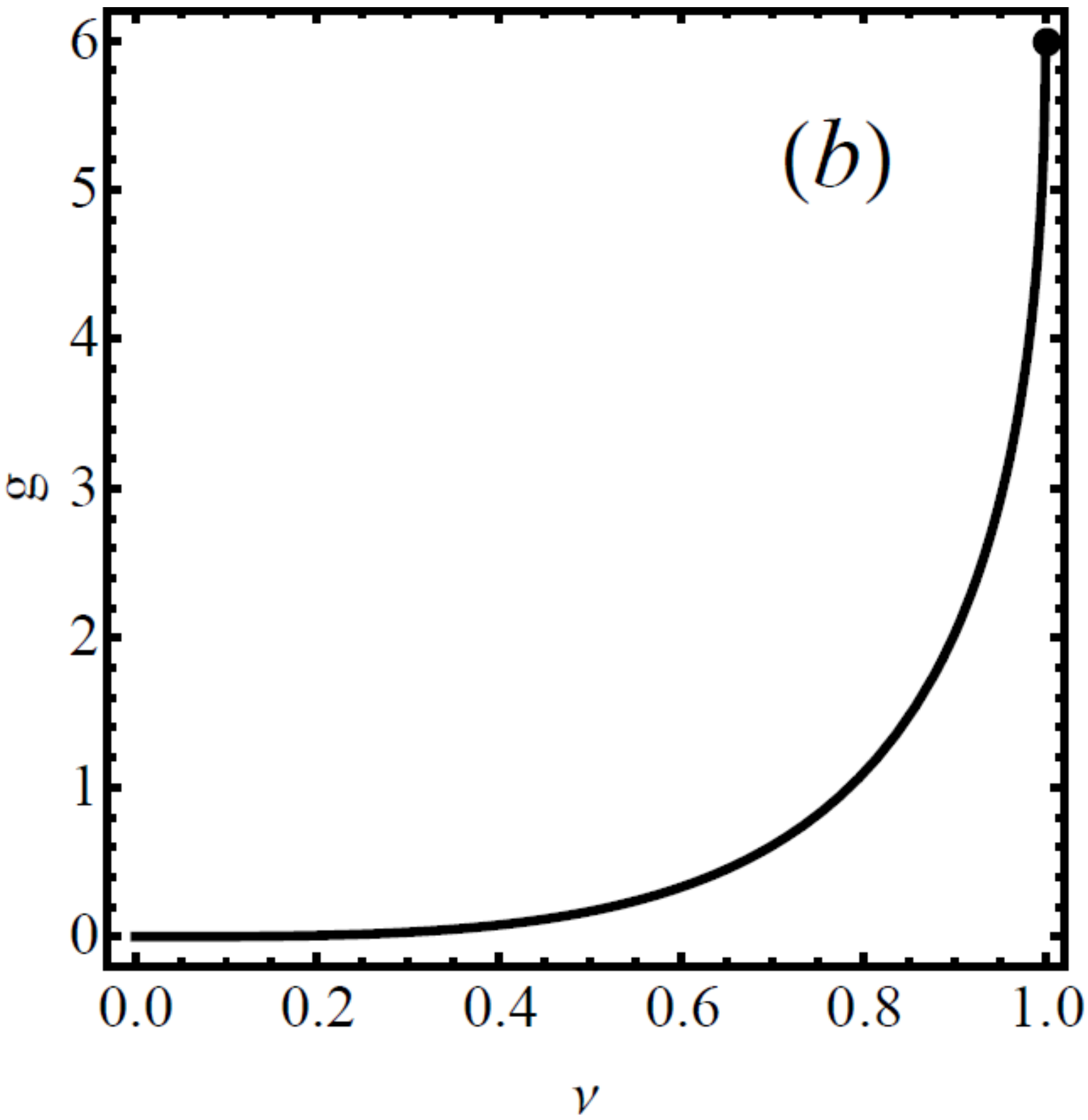}
	\end{tabular}	
\caption{(a): The stationary optimal profile $\rho(x)$ for the ZRP, given by Eqs.~(\ref{123}) and (\ref{zrpparameter}) for $\nu=0.7$. The two fat dots mark the interval boundaries $|x|=1$. The edges of compact support are at $|x|=x_0(\nu=0.7)= 2.76998\dots$.
(b): The rescaled action $g(\nu)$, Eq.~(\ref{szrp}). The survival limit $\nu=1$ is marked by the fat dot. }
\label{szrpfig}		
\end{figure}

The resulting profile of $\rho(x)/n$ is shown in Fig.~\ref{szrpfig}.
Inserting  $\rho(x)$ given by (\ref{123}) and (\ref{zrpparameter}) into Eq.~(\ref{sq}), we finally obtain
\begin{subequations}
\begin{align}
\label{actzrp}
&-\ln\mathcal P(\nu,N) \simeq \frac{\Gamma N^2T}{4l^3} g(\nu), \\
&g(\nu)=\frac{3}{4}\left[27-18\nu-\nu^2-\sqrt{1-\nu}\left(9-\nu\right)^{3/2}\right].
\label{szrp}
\end{align}
\end{subequations}
This function $g(\nu)$ is shown in Fig.~\ref{szrpfig}. As to be expected, $g(\nu=0)=0$. The maximum value $g(\nu=1)=6$ corresponds to survival of all the particles. It is in agreement with Ref.~\cite{surv}, which derived an implicit expression for the survival probability of a diffusive gas with arbitrary $D(\rho)$ and $\sigma(\rho)$.

A solution with compact support, like the one described by (\ref{123}), is also a solution for a \emph{finite} system, as long as the former  fits into the latter. If it does not, the solution must adapt to the boundary conditions and change. As we will show shortly, this leads to a dynamical phase transition. But let us start with a brief discussion of general aspects of the occupation statistics in finite systems.

\section{Occupation statistics on a ring}
\label{ring}

Now suppose that $N$ particles are released on a ring of length $2L$. We set $|x|<L$ and study the time-averaged occupation (\ref{cons}) of a subinterval $|x|<l<L$.  At long times the \emph{expected} particle number density, governed by a deterministic diffusion equation,  approaches the constant value $N/(2L)$. Thus the average occupation of the subinterval is $\bar{\nu}=l/L=1/\mathcal L$, where the rescaled ring length $\mathcal L= L/l>1$ is an additional dimensionless parameter of the problem. Fluctuations  can lead to overpopulation, $\bar{\nu}\leq\nu\leq1$, or underpopulation, $0\leq\nu\leq\bar{\nu}$. The distribution of $\nu$ can be obtained by using the stationary MFT formalism of Sec.~\ref{fluc} with slight modifications. Now Eqs.~(\ref{lagrangein}) and~(\ref{lagrangeout}) are to be solved on the interval $[-\mathcal L,\mathcal L]$ with periodic boundary conditions.  Due to the symmetry of our coordinate system, the derivative $u'(x)$ must vanish at $|x|=\mathcal L$. In addition, the integrals in Eq.~(\ref{mass2}) and~(\ref{su}) should be from $[-\mathcal L,\mathcal L]$.  Doing the same rescaling as the one leading to (\ref{lag2}), we obtain
\begin{equation}
\label{lag22}
s\left(\nu,N,L;l\right)=\frac{1}{l}\tilde{s}\left(\nu,n,\mathcal L\right).
\end{equation}
Since $\nu$ is the fraction of the particles occupying the interval of length $2l$, the complementary $1-\nu$  fraction of the particles occupies the complementary interval of length $2\left(L-l\right)$. Hence  the action satisfies a duality relation
\begin{eqnarray}
s\left(\nu,N,L;l\right)= s\left(1-\nu,N,L;L-l\right).
\label{dual}
\end{eqnarray}
The same argument leads to a duality relation for the optimal density profile:
\begin{eqnarray}
\rho\left(x,\nu,n,\mathcal L\right)=\rho\left(\frac{\mathcal L-x}{\mathcal L-1},1-\nu,\frac{n}{\mathcal L-1},\frac{\mathcal L}{\mathcal L-1}\right)\label{dual2}.
\end{eqnarray}
Because of the duality it suffices to solve the problem only for overpopulation fluctuations with parameters
$$
\left(1/\mathcal L\leq\nu\leq1,n,\mathcal L\right).
$$
The solution for underpopulation fluctuations is then obtained from the overpopulation solution with the parameters $\left(1/\mathcal {L}^{\prime}\leq\nu,n^{\prime},\mathcal L^{\prime}\right)$ for $\mathcal L^{\prime}= \mathcal L/\left(\mathcal L-1\right)$, and $n^{\prime}=n/\left(\mathcal L-1\right)$ by using Eqs.~(\ref{dual}) and~(\ref{dual2}). In the survival limit $\nu=1$ the particles do not leave the subinterval, and the solution coincides with that on the infinite line. Here the duality gives the solution for the void  limit, $\nu=0$.

Now let us consider the ZRP where the interparticle interactions lead to a phase transition. For completeness, we consider in Appendix \ref{ringrws} the non-interacting RWs on a ring, where no phase transition occurs.

\subsection{ZRP on a ring: a dynamical phase transition}

For the lattice gases, which admit stationary optimal solutions with compact support, not only the survival limit $\nu=1$, but a whole range of overpopulation fluctuations $\bar{\nu}\leq\nu_c\leq\nu$ (with a critical value $\nu_c$ we will soon determine) is described by the infinite-system solution. For simplicity, we again consider the ZRP with $\alpha=1$. Here the solution (\ref{123}) with compact support also applies for a finite ring as long as the ring size $\mathcal L$ is larger than the size of support $x_0\left(\nu\right)$ (\ref{zrpparameter}). This condition defines a critical ring size, $\mathcal L_c\left(\nu\right)= x_0\left(\nu\right)$. By inverting the relation (\ref{zrpparameter}) we see that, for a given ring size $\mathcal L$, the $\mathcal L$-independent solution (\ref{123}) with compact support solves the ring problem when $\nu$ is equal to or larger than the critical value $\bar{\nu}<\nu_{c1}<1$, which is given by
\begin{equation}
\nu_{c1}\left(\mathcal L\right)=\frac{3\mathcal L-1}{\mathcal L\left(\mathcal L+1\right)}.
\end{equation}
When the occupation fraction is at the critical value $\nu=\nu_{c1}(\mathcal L)$ [or for a ring size at the critical value $\mathcal L=\mathcal L_c(\nu)$], the solution \eqref{123} vanishes at the ring's boundary. Upon a further decrease of the occupation fraction (alternatively for a further decrease in the ring size), the solution with compact support of Eq.~(\ref{3z}) crosses over to a solution, the support of which is the whole ring. This solution has a positive minimal value $\rho_{\text{min}}$ at the boundary $|x|=\mathcal L$:
\begin{equation}
\label{u1zr}
\frac{\rho(x)}{n}=
\begin{cases}
\rho_{\text{max}}-\lambda_1^2x^2                & |x|<1, \\
\lambda_2^2\left(|x|-\mathcal L\right)^2 +\rho_{\text{min}}                 & 1<|x|<\mathcal L.
\end{cases}
\end{equation}
The  constants $\rho_{\text{max}}$, $\rho_{\text{min}}$, $\lambda_1$ and $\lambda_2$ are determined from the continuity of $\rho$ and $\rho_x$  at $|x|=1$ and the constraints (\ref{cons2}) and (\ref{mass2}), and we  obtain \cite{fornu}:
\begin{eqnarray}
\lambda_1^2&=&\left(\mathcal L-1\right)\lambda_2^2=\frac{3}{2\left(\mathcal L-1\right)}\left(\nu-\frac{1}{\mathcal L}\right),\nonumber\\
\rho_{\text{max}}
&=&\frac{2\mathcal L-1}{2\left(\mathcal L-1\right)}\left(\nu-\nu_{c2}\right),\\
\rho_{\text{min}}&=&\frac{\mathcal L+1}{2\left(\mathcal L-1\right)}\left(\nu_{c1}-\nu\right),\nonumber
\label{u3zr}
\end{eqnarray}
where we have denoted
\begin{equation}
\label{n2}
\nu_{c2}\left(\mathcal L\right)=\frac{1}{\mathcal L\left(2\mathcal L-1\right)}\leq\bar{\nu}.
\end{equation}

\begin{figure}[h]
	\begin{tabular}{ll}
		\includegraphics[width=0.23\textwidth,clip=]{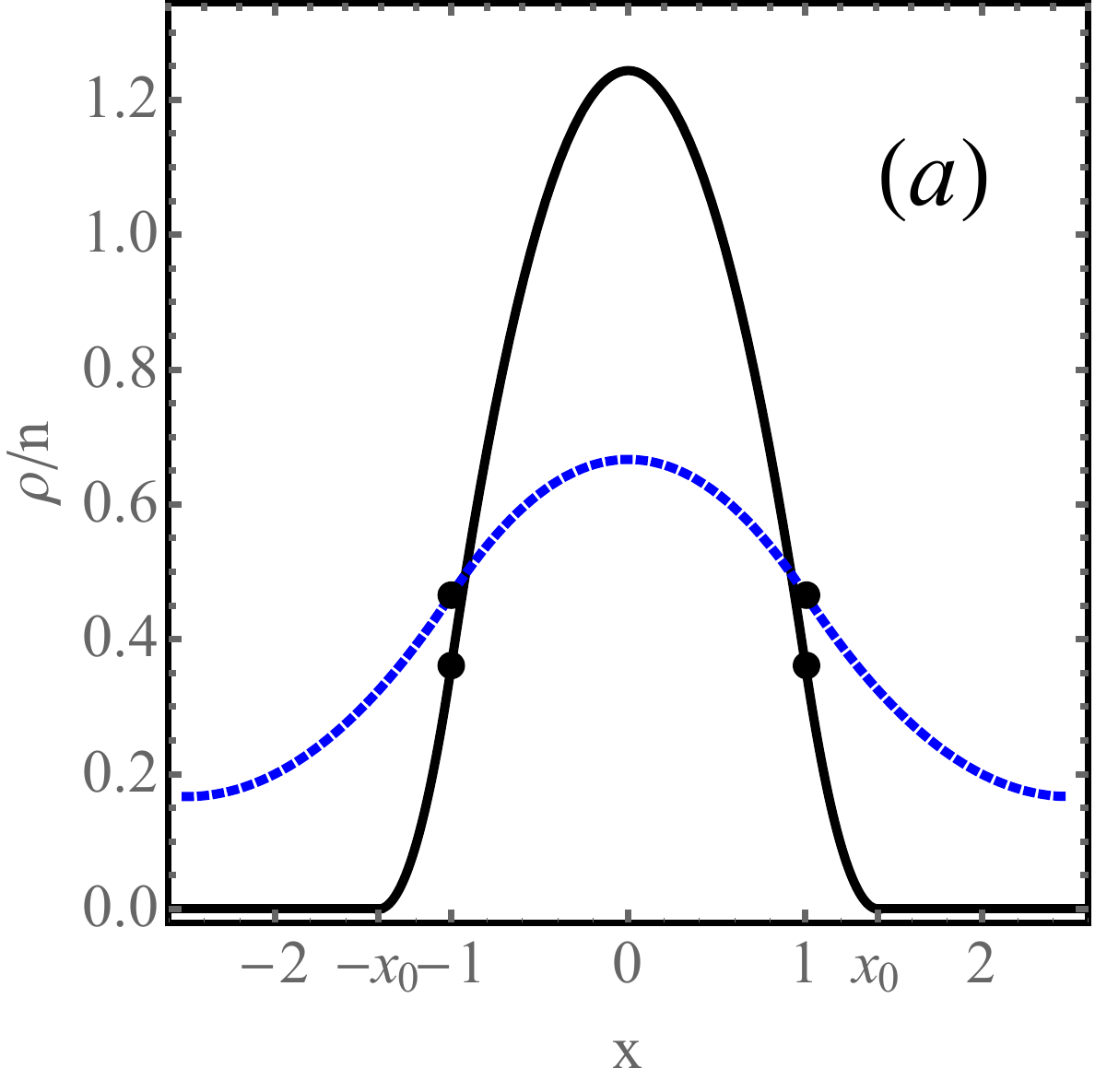}&
		\includegraphics[width=0.23\textwidth,clip=]{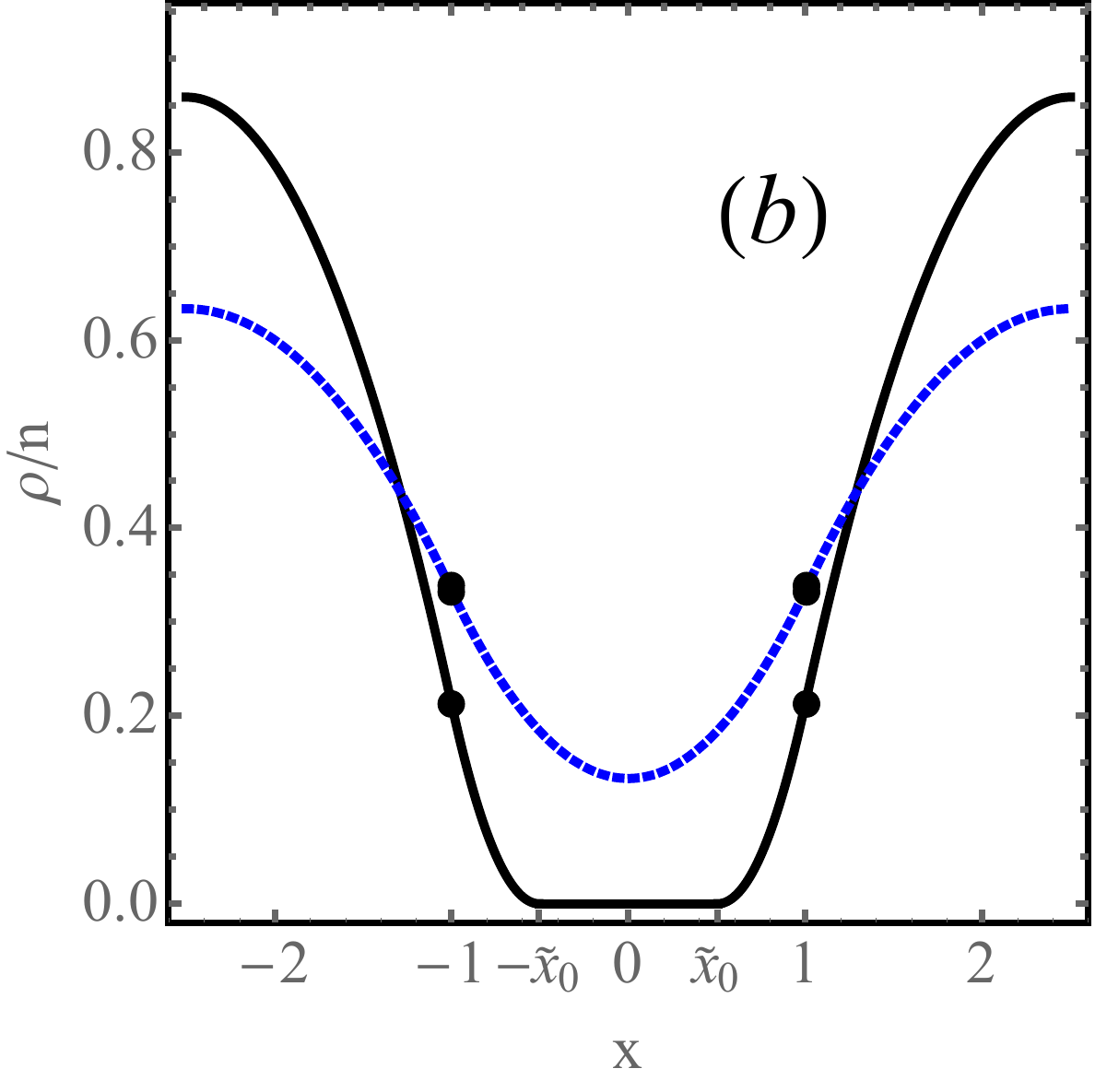}
	\end{tabular}	
	\caption{The stationary optimal profile $\rho(x)$  for the ZRP on a ring of the rescaled size $\mathcal L=2.5$ for the overpopulation fluctuations (a) and underpopulation fluctuations (b), see the main text for details. The solid lines describe the solutions with compact support for $\nu=0.95>\nu_{c1}=0.74\dots$ (a) and $\nu=0.035<\nu_{c2}=0.1$ (b). The boundaries of compact support are at $x_0=1.41\dots$ (a) and $\tilde{x}_0=0.50\dots$ (b). The dotted line corresponds to the extended solution (\ref{u1zr}), obtained for $\nu=0.6<\nu_{c1}$ (a) and $\nu=0.2>\nu_{c2}$ (b). The two fat dots mark the boundaries between the interior, $|x|<1$, and the exterior, $|x|>1$, regions of these solutions.}
\label{qzrpring}
\end{figure}

At $\nu=\nu_{c1}(\mathcal L)$ we obtain $\rho_{\text{min}}=0$, and the extended solution (\ref{u1zr}) coincides with the solution \eqref{123} with compact support. The duality relation \eqref{dual2} helps us obtain the solution in the case of underpopulation, $0\leq\nu\leq\bar{\nu}$. Indeed, the extended solution (\ref{u3zr}) holds down to the critical occupation fraction $\nu=\nu_2\left(\mathcal L\right)$. At $0\leq\nu\leq \nu_{c2}\left(\mathcal L\right)$ the solution crosses over to a compact underpopulation solution, which is dual to the one given by (\ref{123}); the latter now fits into the ring). These solutions are shown in Fig.~\ref{qzrpring} for overpopulation (a) and underpopulation (b) fluctuations.

The transitions between the two types of solutions result in a non-analyticity of the action  at $\nu_{c1}$ and $\nu_{c2}$. For $\bar{\nu}\leq\nu\leq\nu_{c1}$ the action, evaluated over the extended solution (\ref{u1zr}), yields a Gaussian distribution:
\begin{equation}
-\ln\mathcal P(\nu,N,L)\simeq\frac{3\Gamma TN^2L}{2l^2\left(L-l\right)^2}\left(\nu-\bar{\nu}\right)^2.\label{szrpr2}
\end{equation}
This result, alongside with Eq.~(\ref{actzrp}), yields the action in the overpopulation region $\bar{\nu}\leq\nu\leq 1$. The underpopulation region $0\leq\nu\leq\bar{\nu}$ is obtained from the duality relation (\ref{dual}). Specializing it to the ZRP scaling
\begin{equation}
-\ln\mathcal P(\nu,N,L)\simeq\frac{\Gamma TN^2}{4l^3}g^{\text{ring}}\left(\nu,\mathcal L\right),
\end{equation}
we see that $g^{\text{ring}}\left(\nu,\mathcal L\right)$ obeys the duality relation
\begin{equation}
g^{\text{ring}}(\nu,\mathcal L)=
\frac{g^{\text{ring}}(1-\nu,\frac{\mathcal L}{\mathcal L -1})}{\left(\mathcal L -1\right)^3}
\end{equation}
and  we obtain $g^{\text{ring}}$ over the entire parameter range:
\begin{numcases}
{\!\!g^{\text{ring}}\left(\nu,\mathcal L\right)=}
g^{\text{line}}\left(\nu\right),\!\!\!\!\!\!\! & $\nu_{c1}\leq\nu\leq 1$, \label{high}\\
\frac{6\mathcal L\left(\nu-\bar{\nu}\right)^2 }{\left(\mathcal L-1\right)^2},\!\!\!\!\!\!\!& $\nu_{c2}\leq\nu\leq\nu_{c1}$, \label{gauss}\\
\frac{g^{\text{line}}\left(1-\nu\right) }{\left(\mathcal L-1\right)^3},\!\!\!\!\!\!\!& $0\leq\nu\leq \nu_{c2}$, \label{actionzrpring}
\end{numcases}
where $g^{\text{line}}$ is the corresponding function for the infinite line, Eq.~(\ref{szrp}).

\begin{figure}[h]
	\begin{tabular}{ll}
		\includegraphics[width=0.23\textwidth,clip=]{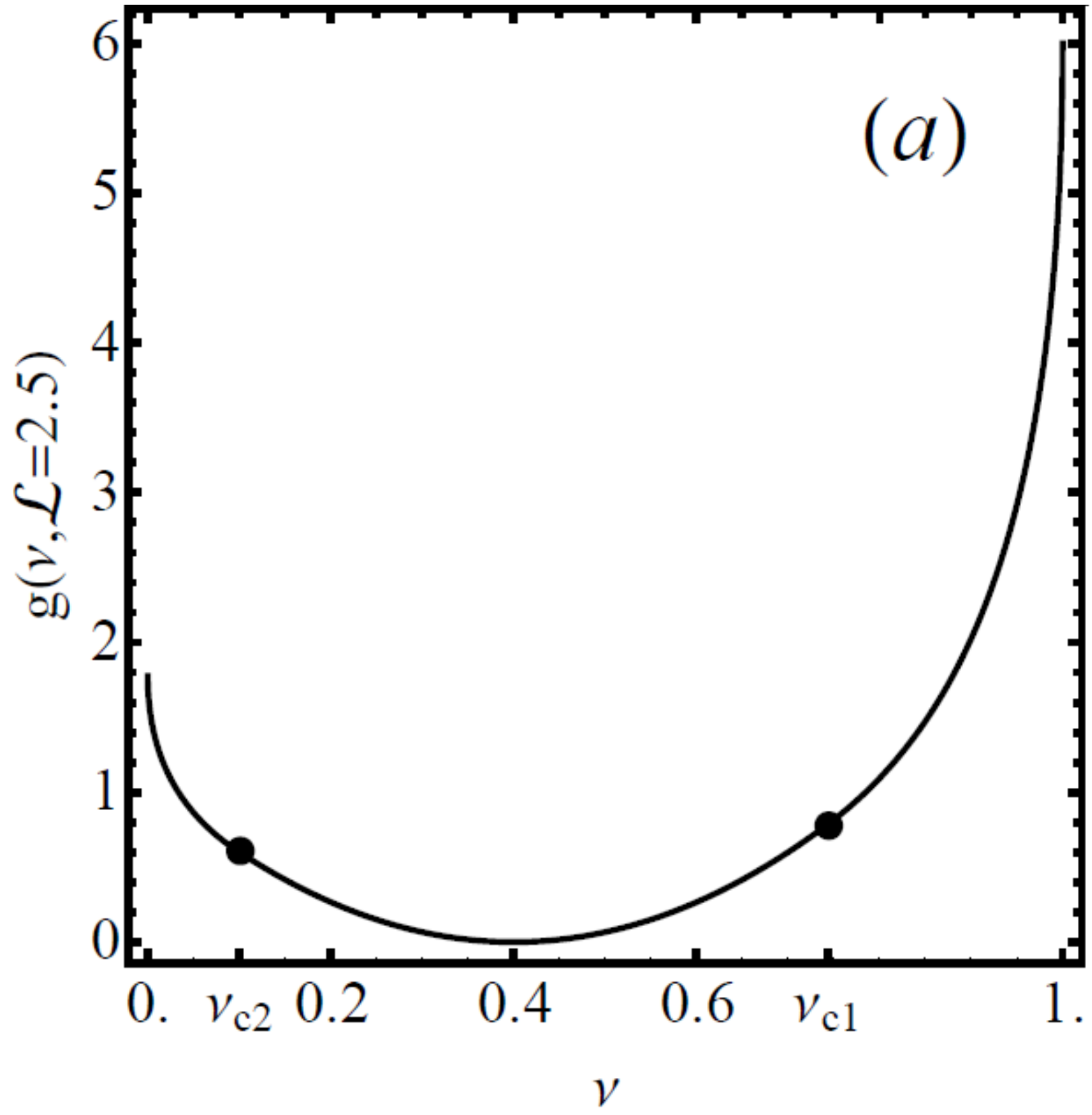}
		&
		\includegraphics[width=0.237\textwidth,clip=]{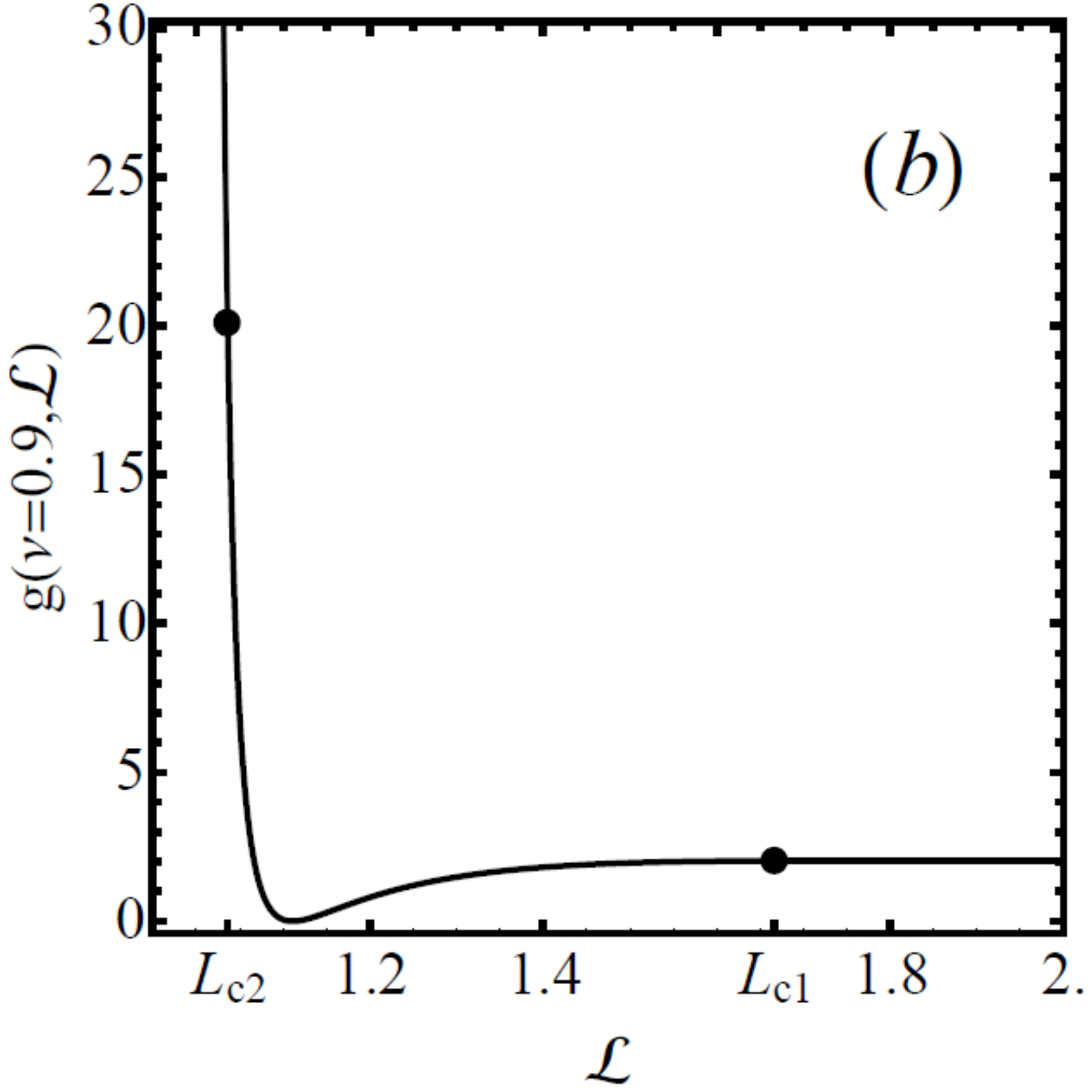}
	\end{tabular}	
	\caption{The action $g^{\text{ring}}(\nu,\mathcal L)$ for the ZRP on a ring, described by Eqs.~(\ref{high})-(\ref{actionzrpring}). (a): $g^{\text{ring}}(\nu,\mathcal L =2.5)$ as a function of the occupation fraction $\nu$ for a ring of fixed size $\mathcal L=2.5$. (b):  $g^{\text{ring}}(\nu=0.9,\mathcal L)$ as a function of the $\mathcal L$ for a fixed $\nu=0.9$. The fat dots mark the points of the second-order  phase transition at $\nu_{c1}=0.74\dots$ and $\nu_{c2}=0.1$ (a), and $\mathcal L_{c1}=1.66\dots$ and  $\mathcal L_{c2}=1.03\dots$ (b).
	}\label{szrpring}
\end{figure}

Figure~\ref{szrpring} shows $g^{\text{ring}}\left(\nu,\mathcal L\right)$ as a function of $\nu$ for a fixed $\mathcal L=2.5$ (a) and as function of $\mathcal L$ for a fixed $\nu=0.9$. The latter dependence is non-monotonic.  For $\mathcal L > \mathcal L_{c1}(\nu=0.9)=1.66\dots$ the solution with compact support fits into the ring, and $g$ is independent of the ring size $\mathcal L$. The corresponding critical ring size $\mathcal L < \mathcal {L}_{c2 }(\nu=0.9)=1.03\dots$ is obtained from the duality relation (\ref{dual}). Below  $\mathcal {L}_{c2 }$ an underpopulation solution with  compact support fits into the ring. In this region of parameters the action depends on $\mathcal L$, because the size of the populated region depends on $\mathcal L$. The action diverges when $\mathcal L$ approaches the minimal size, $g(\nu,\mathcal L \rightarrow 1)\sim \left(\mathcal L -1 \right)^{-3}$ [see Eq.~(\ref{actionzrpring})], as the system attempts to accommodate a finite number of particles inside a segment of a vanishing length. Finally, for an intermediate value $\mathcal L=1/\nu=10/9$, the action vanishes, because in this case the occupation fraction $\nu=0.9$ is achieved when the density profile is flat (and deterministic).

\begin{figure}
	\includegraphics[width=0.35\textwidth,clip=]{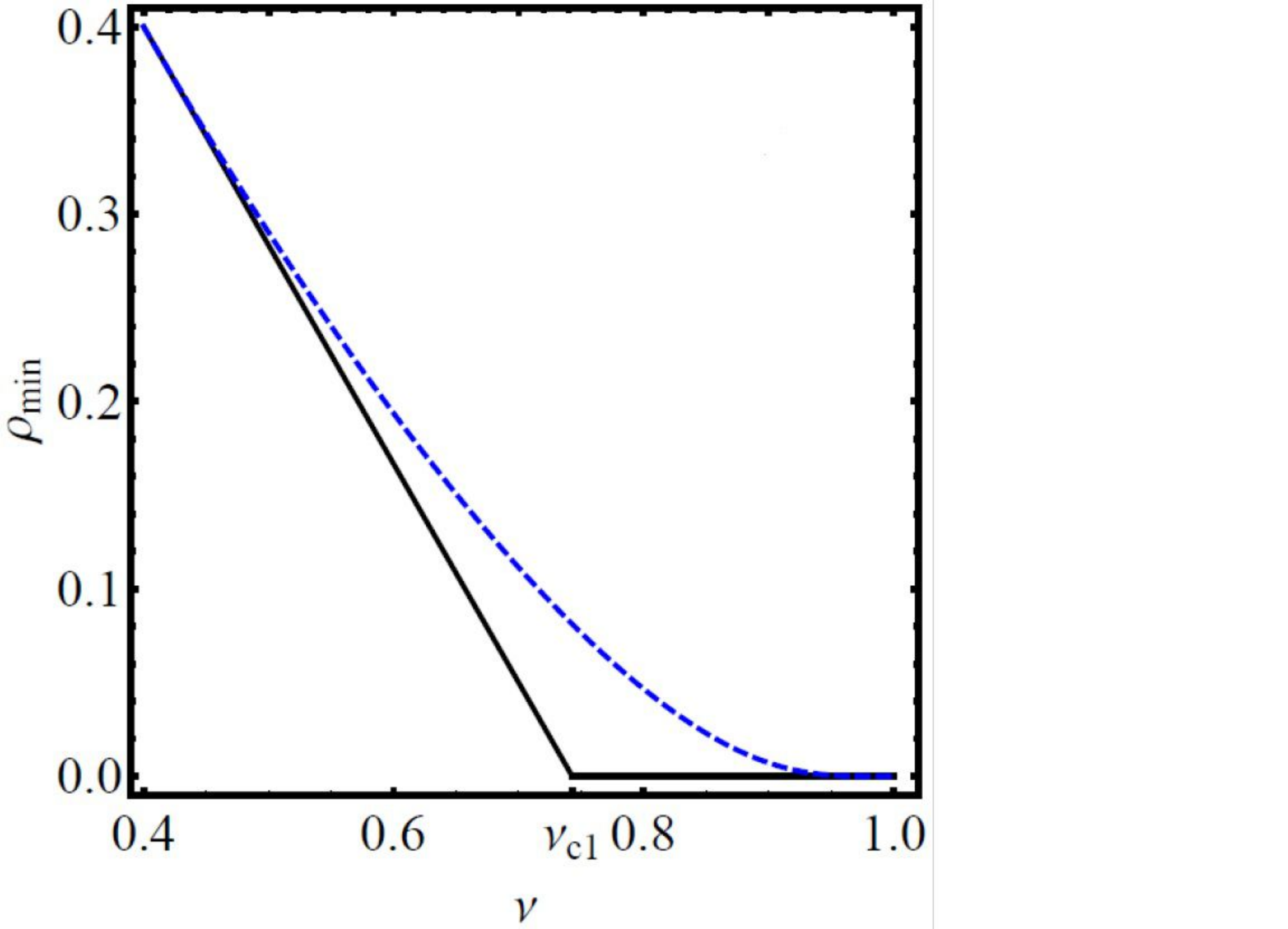}
	\caption{The minimum density $\rho_{\text{min}}$  as a function of the occupation fraction $\nu$ for overpopulation fluctuations $\bar{\nu}\leq\nu\leq1$ in a ring of size $\mathcal L=2.5$   (solid line). Note the sharp transition at $\nu_{c1}=0.74\dots$. For comparison, the dashed line shows the same quantity for the RWs (see Appendix \ref{ringrws}), where there is no phase transition.}
	\label{Fig:order}
\end{figure}

The transitions between the Gaussian region of the action (\ref{gauss}) and the non-Gaussian parts (\ref{high}) and (\ref{actionzrpring}) are accompanied by discontinuities of the second derivative of the action. Such discontinuities are called dynamical phase transition.
A natural choice for an order parameter to characterize this dynamical phase transition is the minimum density value $\rho_{\text{min}}$, which decreases linearly below the critical point  $\nu_{c1}$ and vanishes at and above the critical point (Fig.~\ref{Fig:order}). This behavior is in a stark contrast to the square-root behavior of the order parameter in the usual Landau-type second order phase transition \cite{land}. Indeed, the origin of the second-order phase transition in the ZRP is  the non-negativity constraint (\ref{positive}), rather than a symmetry breaking of the optimal profile. In the absence of the non-negativity constraint, the minimization problem always has a \emph{unique} solution, given by Eq.~(\ref{u1zr}). Within the interval $\nu_{c2}\leq\nu\leq\nu_{c1}$, where it does not violate Eq.~(\ref{positive}), this solution is also the solution to the constrained problem. Outside of  this interval of $\nu$, this solution is forbidden, as it enters the forbidden zone of $\rho<0$; the minimization procedure  must incorporate the non-negativity constraint, via a tangent construction, leading to the solution \eqref{123} with compact support. The transition between the unconstrained minimizing solution (\ref{u1zr}) and the one-sided solution \eqref{123}  lies at the origin of the second-order phase transition.

A similar dynamical phase transition should appear in all diffusive lattice gas models, whose stationary optimal solution on the infinite line has a compact support. In particular, this property is shared by all ZRP models, see Sec.~\ref{fluc}, with $\alpha>0$.

\section{Summary and Discussion}\label{disc}

Inter-particle interactions can strongly affect  the long-time occupation statistics of an ensemble of diffusing particles. Here we employed the macroscopic fluctuation theory (MFT) to uncover some of these effects.
As we have seen, the MFT is also very useful in the absence of interactions, as it provides an insightful information about the most likely history of the particle system, conditioned on a specified occupation fraction.

The occupation statistics of interacting particles depends in a non-trivial way on the total number of particles. A more surprising effect is the second-order dynamical phase transition, where  the rate function $s(\nu,N)$ (\ref{gas}) is non-analytic with respect to the occupation fraction $\nu$. This transition appears, in finite systems, in a whole class of interacting particle models for which the optimal stationary density profile $\rho(x)$ has compact support. A simple example of such models is the zero range process where the hopping rate  to the neighboring sites increases faster than linearly with the number of particles on the departure site.

A dynamical phase transition of a very different nature was recently uncovered for the occupation statistics of a single Brownian particle, driven by external force \cite{Hugo18,Hugo16b}, see also earlier work \cite{Nieuwenhuizen, Grassberger}.  The dynamical phase transition, that we found here, appears for an equilibrium system without any external driving, and it is a consequence of interactions between the particles. It would be interesting to find out whether different types of interactions, encoded in the density dependence of the system's diffusivity and mobility, bring about additional types of singularities of the rate function [see \textit{e.g.} the footnote preceding Eq.~(\ref{mass})].

Our findings strongly rely on the stationarity assumption, that is, on the additivity principle, see Sec. \ref{fluc}. For non-interacting RWs, the additivity principle can be verified both via the MFT approach (by solving the full time-dependent MFT problem), and via the exact single-particle approach. For the SSEP, the additivity principle was established only in the survival limit $\nu=1$ \cite{surv}. If for some $D(\rho)$ and $\sigma(\rho)$ the additivity principle breaks down at a critical value of $\nu$, one will observe a yet another type of dynamical phase transition \cite{ber,main2,hurtado,main,phase}.  This is a very interesting direction to explore.

Our MFT formalism for the occupation statistics can be extended to higher dimensions and to more complicated geometries. The limiting case of the survival, $\nu=1$, in higher dimensions has already been studied with the MFT \cite{surv,MVK}, where the long-time statistics come from the additivity principle. Assuming that the additivity principle holds for the entire range of occupation fraction $0\leq\nu\leq1$, we can immediately predict the exponential law (\ref{gas}) in higher dimensions as well.

It would be very interesting to measure the occupation statistics of ensembles of particles in experiment. One possible method is fluorescence correlation spectroscopy (FCS) \cite{flor}. In the basic FCS setup, a laser beam is focused into an observation region inside a suspension containing fluorescent Brownian particles. The particles are at the focal volume fluoresce, and the emitted light is registered. The emitted light power fluctuates due to fluctuations in the instantaneous number of fluorescent particles in the observation region, so the statistics of the total emitted light over the entire measurement time should be described by the occupation fraction statistics. For sufficiently high densities of the fluorescent particles, inter-particle interactions should lead to deviations from the statistics (\ref{singlerw}) based on the single-particle calculations.

Finally, this work has established an explicit mathematical equivalence between the stationary MFT formalism for non-interacting particles and two large deviation theories:  the level 2 formalism and the Donsker-Varadhan formalism. In particular, this equivalence provides a simple relation between the optimal particle number density in the MFT formalism and the single-particle probability distribution of the conditioned process, as discussed in Sec.~\ref{RWs}. To our knowledge, this equivalence has not yet been addressed in detail. It would be interesting to understand it better, and find out whether it can be  extended beyond the long-time limit, or for other large-deviation problems which do not necessarily involve empirical measures.

\section*{ACKNOWLEDGMENTS}

We thank Sidney Redner, Naftali Smith and Hugo Touchette for useful discussions. TA and BM acknowledge support from the Israel Science Foundation (Grant No. 807/16).

\appendix
\section{Dilute limit of the SSEP}\label{rwsapend}

 In the limit $k\rightarrow0$ we can replace $\text{sn}(...,k)\rightarrow\sin(...)$, $\text{cn}(...,k)\rightarrow\cos(...)$, and $\text{dn}(...,k)\rightarrow1$. Using these asymptotics in Eqs.~(\ref{l2})--(\ref{qrho}), we obtain, after some algebra, the following leading-order expressions at $k\ll1$:
 \begin{eqnarray}
 \lambda_2 &\simeq&\lambda_1\tan\lambda_1,\label{paral1}\\
 A&\simeq&\lambda_1\tan\lambda_1-\ln\left(\frac{2}{k\cos\lambda_1}\right),\label{paral2}\\
 n&\simeq&k^2\frac{\cot\lambda_1+\lambda_1}{2\lambda_1}\label{paral3},\\
 \nu&\simeq&\frac{\sin\lambda_1\cos\lambda_1+\lambda_1}{\cot\lambda_1+\lambda_1}.\label{paral4}
 \end{eqnarray}
The optimal profile (\ref{qssep}) can be approximated as
\begin{equation}
\label{u2ap}
\rho(x)=
\begin{cases}
k^2\cos^2(\lambda_1x)   &|x|<1,\\
k^2\cos^2\lambda_1 \times e^{2\lambda_2(1-|x|)}  & |x|>1.
\end{cases}
\end{equation}
This optimal profile together with Eqs.~(\ref{paral1})--(\ref{paral4}) exactly reproduce our results (\ref{qrws})--(\ref{para3}) for the RWs. Finally, the action (\ref{sssep}) in this limit becomes
 \begin{equation}
 g(\nu,n)\simeq\lambda_1^2k^2.
 \end{equation}
This result, together with Eqs.~(\ref{paral3}) and (\ref{paral4}), reproduces Eqs. (\ref{srws}) and (\ref{para3}) leading to Eq.~(\ref{rs}).

\section{RWs on a ring}\label{ringrws}

\subsection{General}

\begin{figure}[h]
	\begin{tabular}{ll}
		\includegraphics[width=0.23\textwidth,clip=]{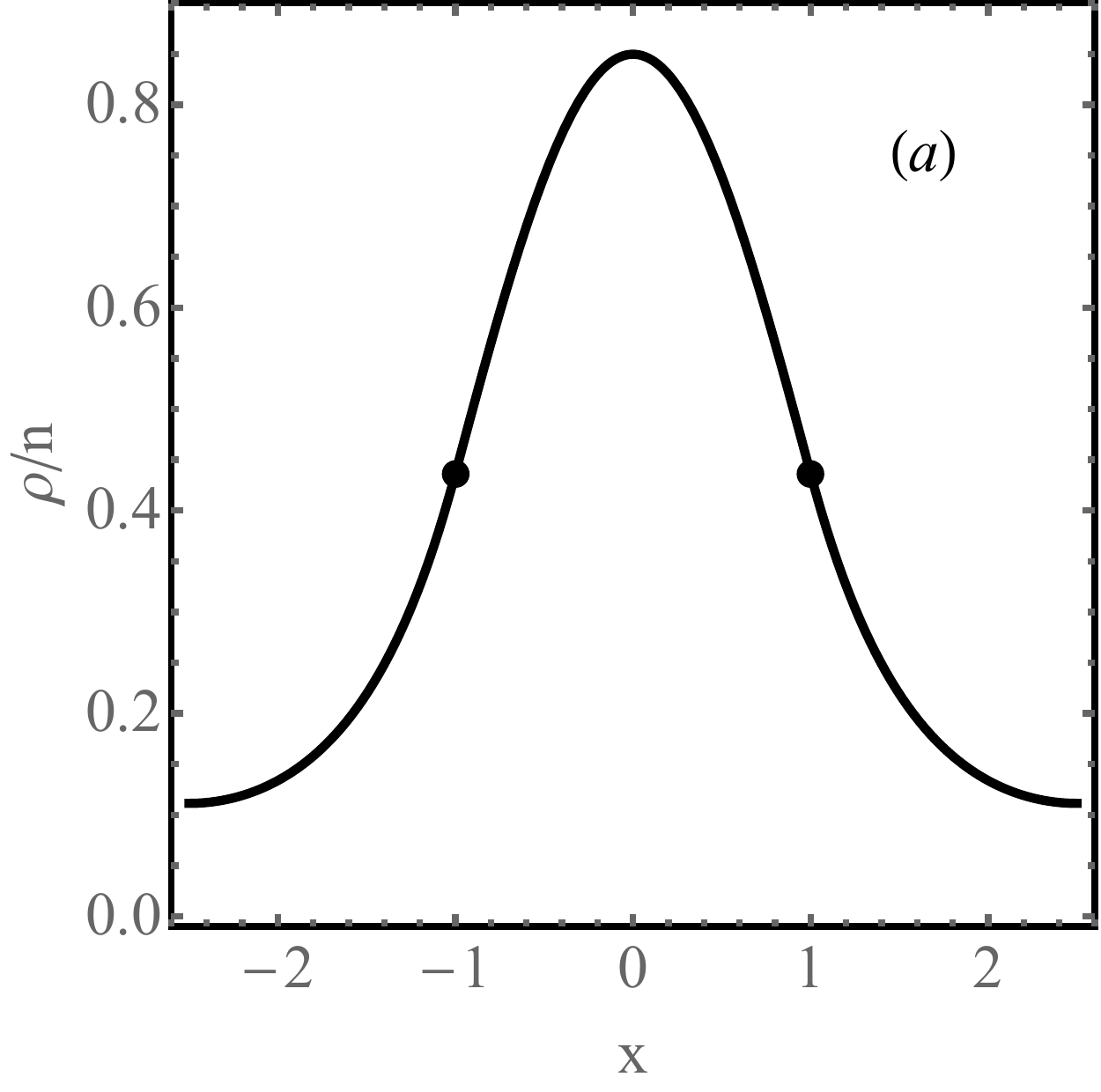}
		&
		\includegraphics[width=0.22\textwidth,clip=]{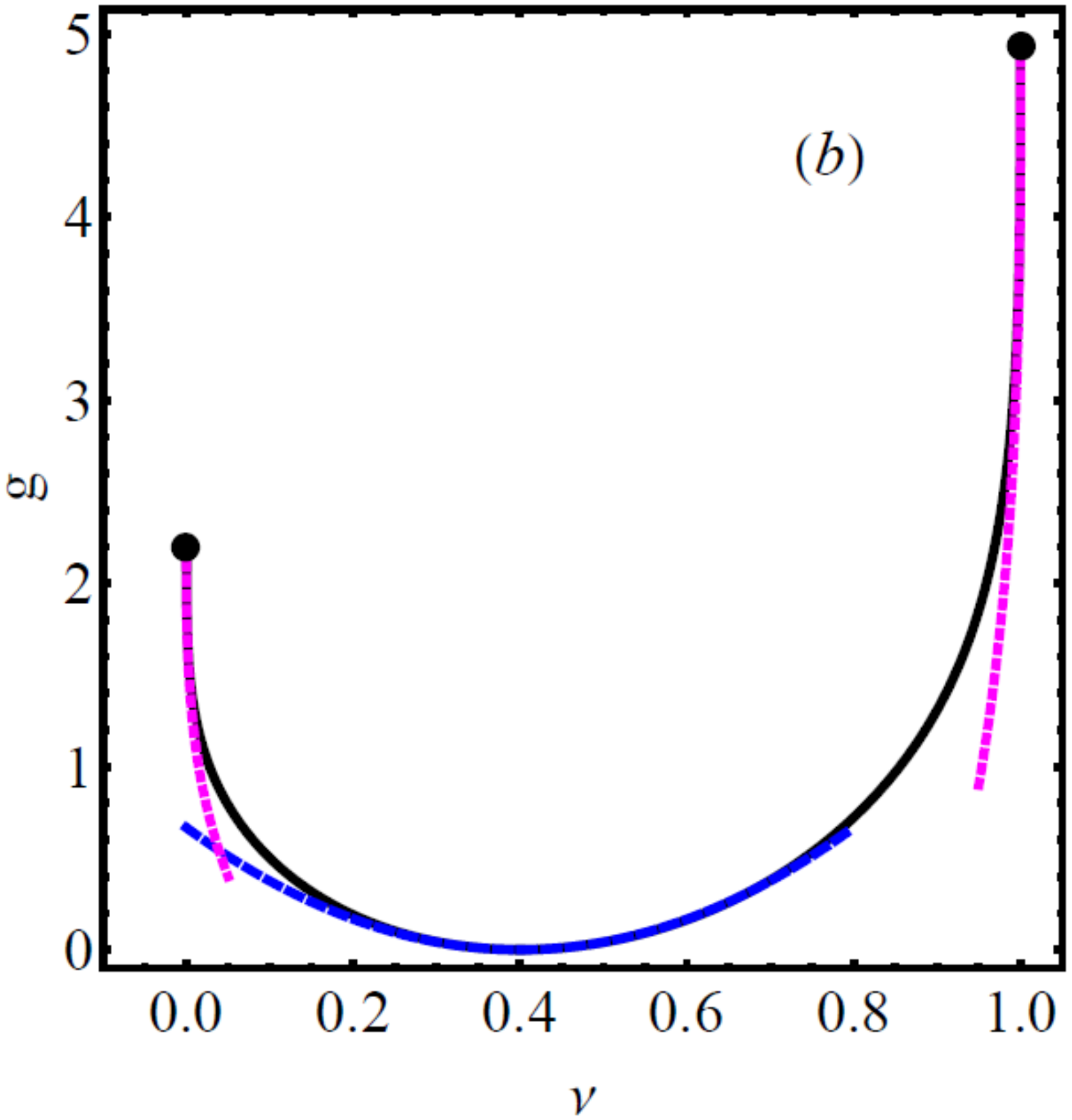}
	\end{tabular}	
	\caption{(a): The stationary optimal profile $\rho(x)$ for the RWs inside a ring of length $\mathcal L=2.5$, given by Eqs.~(\ref{qrwsring}) and (\ref{ab12}) for $\nu=0.7$. The two fat dots mark the boundary between the interior $|x|<1$ and the exterior $|x|>1$ regions. (b): The rescaled action $g(\nu,\mathcal L=2.5)$, Eq.~(\ref{srwsring}). The extreme limits of survival, $g(\nu=1,\mathcal L=2.5)=\pi^2/2$, and void, $g(\nu=0,\mathcal L=2.5)=\pi^2/2\left(\mathcal L-1\right)^2$, are marked by fat dots. The dashed line is the Gaussian asymptotic (\ref{lowring}), and the dotted lines are the asymptotics of survival (\ref{highring}) and void (\ref{highring2}).}
	\label{qrwsringfig}		
\end{figure}

We start by considering overpopulation fluctuations, $\bar{\nu}\leq\nu$, and solve the minimization problem (\ref{helm}) on the ring. A symmetric  periodic solution can be written as
\begin{equation}
\label{u2}
\frac{u}{2\sqrt{nD_0}}=
\begin{cases}
\sqrt{\nu} A\cos(\lambda_1x)    &|x|<1,\\
\sqrt{\frac{1-\nu}{\mathcal L-1}}B\cosh X  & 1<|x|<l,
\end{cases}
\end{equation}
where we have shortly written
\begin{equation}
\label{X:def}
X = \frac{\lambda_2\left(\mathcal L-|x|\right)}{\mathcal L-1}
\end{equation}
and rescaled the Lagrange multiplier $\lambda_2\rightarrow\left(\mathcal L-1\right)\lambda_2$.
Imposing the constraints~(\ref{cons2}) and~(\ref{mass2}), we can express $A$ and $B$ in terms of $\lambda_1$ and $\lambda_2$:
\begin{eqnarray}
A^2\left[\frac{\sin\left(2\lambda_1\right)}{2\lambda_1}+1\right]&=&1,\\\label{c1}
B^2\left[\frac{\sinh\left(2\lambda_2\right)}{2\lambda_2}+1\right]&=&1.\label{c2}
\end{eqnarray}
Imposing the matching conditions at $|x|=1$ we obtain
\begin{eqnarray}
\frac{A\cos\lambda_1}{B\cosh\lambda_2}&=&\sqrt{\frac{1-\nu}{\mathcal (L-1)\nu}},\\\label{c3}
\frac{\lambda_1\tan\lambda_1}{\lambda_2\tanh\lambda_2}&=&\frac{1}{\mathcal L-1}.\label{c4}
\end{eqnarray}
The solution can be obtained in a \emph{double} parametric form, where the parameters $0\leq\lambda_1\leq\pi/2$ and $\lambda_2\geq 0$ corresponds to $0\leq\nu\leq 1$ and $\mathcal L\geq 1$. The optimal profile  $\rho(x)=u^2(x)/2D_0$ reads
\begin{equation}
\label{qrwsring}
\frac{\rho\left(x\right)}{n}=
\begin{cases}
a(\lambda_1,\lambda_2)\cos^2(\lambda_1x) & |x|<1, \\
b(\lambda_1,\lambda_2)\cosh^2 X  & 1<|x|\leq \mathcal L,
\end{cases}
\end{equation}
where
\begin{equation}
\label{ab12}
\begin{split}
a(\lambda_1,\lambda_2)&= \frac{2\nu \lambda_1}{\sin\lambda_1\cos\lambda_1+\lambda_1}\,, \\
b(\lambda_1,\lambda_2)&= \frac{2(1-\nu)\lambda_2}{(\mathcal L-1)\left(\sinh\lambda_2\cosh\lambda_2+\lambda_2\right)}\,,
\end{split}
\end{equation}
and the parameters $\lambda_1(\nu,l)$ and $\lambda_2(\nu,l)$ are given implicitly by Eqs.~(\ref{lr}) and (\ref{nur}). An example of the resulting density profile is shown in Fig.~\ref{qrwsringfig}.

Inserting (\ref{qrwsring}) into (\ref{sq}) yields the action (\ref{actionmain}):
\begin{eqnarray}
-\ln\mathcal P(\nu,N,L)&\simeq&\frac{D_0TN}{2 l^2 }g(\nu,\mathcal L),\label{ringactionrws}
\end{eqnarray}
where $g\left(\nu,\mathcal L\right)$ is given  in a double parametric form by
\begin{eqnarray}
&&g(\nu,\mathcal L)=2\lambda_1^2\nu-\frac{2\left(1-\nu\right)\lambda_1^2\tan^2\lambda_1}{\tanh^2\lambda_2},\label{srwsring}\\
&&\mathcal L= 1+\frac{\lambda_2\tanh\lambda_2}{\lambda_1\tan\lambda_1},\label{lr}\\
&&\nu=\left[1+\frac{\cos^3\lambda_1\sinh\lambda_2\left(\sinh 2\lambda_2+2\lambda_2\right)}{\cosh^3\lambda_2\sin\lambda_1\left(\sin2\lambda_1+2\lambda_1\right)}\right]^{-1}.\label{nur}
\end{eqnarray}
The underpopulation fluctuations follow from these expressions by employing the duality relations (\ref{dual}) and (\ref{dual2}).
Figure~\ref{qrwsringfig} depicts the rescaled action $g$ as a function of $\nu$ at $\mathcal L=2.5$. We now discuss several limiting cases.

The case of infinite ring can be obtained by taking the $\lambda_2\rightarrow\infty$ limit. In the leading order, the hyperbolic cosine in Eq.~(\ref{qrwsring}) turns into the exponent, and the optimal profile (\ref{qrwsring}) coincides with that for the infinite line (\ref{qrws}). Taking the $\lambda_2\rightarrow\infty$ limit in (\ref{srwsring})--(\ref{nur}), we obtain $\mathcal L(\lambda_1,\lambda_2\rightarrow\infty) =\infty$ and reproduce the infinite-line results~(\ref{srws}) and (\ref{para3}).

\subsection{Gaussian fluctuations, $\nu-\bar{\nu}\ll\bar{\nu}$}
This limit corresponds to small deviations of the optimal profile (\ref{qrwsring}) from  the flat equilibrium profile $\bar{\rho}=N/2L$.
Here $\lambda_1\ll1$ and $\lambda_2\ll1$.  Linearizing Eq.~(\ref{lr}) and (\ref{nur}) with respect to $\lambda_1$ and $\lambda_2$, we obtain:
\begin{equation}
\delta \simeq\frac{2}{3}\lambda_2^2,\quad
\mathcal L\simeq1+\left(\frac{\lambda_2}{\lambda_1}\right)^2,
\end{equation}
where $\delta\equiv \nu \mathcal L-1=\nu/\bar{\nu}-1\ll1$.
Solving these equations, we obtain $\lambda_1$ and $\lambda_2$ in the leading order in $\delta\ll1$:
\begin{eqnarray}
\lambda_1^2\simeq \frac{3\delta}{2\left(\mathcal L-1\right)},\quad
\lambda_2^2 \simeq\frac{3}{2}\delta,
\end{eqnarray}
which upon insertion in (\ref{srwsring}) gives a quadratic approximation for the action, see Eq.~(\ref{lowring}) below. This expression suffices for the evaluation of the variance of the occupation fraction fluctuations \cite{KrMe}, see Eq.~(\ref{var}) below. The same result for the variance follows if one uses (\ref{singlerw}) and the single-particle variance obtained in \cite{berez} for a single particle on a segment with reflecting boundary conditions. The close relation between the two settings is obvious within the MFT formalism. Indeed, the optimal density profile in the ring problem has zero derivatives at $x=0$ and $x=L$. As a result, the right half of our periodic solution is also the solution for the system studied in Ref. \cite{berez}.
The action evaluated for the latter system is equal to one half of the action for the ring. The action per particle, however, is the same for both systems.
\bigskip

\subsection{Close to survival, $\nu\rightarrow1$}

The survival limit $\nu=1$ coincides with that of the infinite system and corresponds to $\lambda_1=\pi/2$. The close-to-survival asymptotic of the rescaled action $g$ is obtained by expanding Eqs.~(\ref{srwsring})--(\ref{nur}) near $\lambda_1=\pi/2$. Writing $\epsilon=\frac{\pi}{2}- \lambda_1$ and keeping the leading terms in $\epsilon$ we obtain
\begin{eqnarray}
g&\simeq&\frac{\pi^2}{2}-\pi \epsilon\left(3+\frac{2\lambda_2}{\sinh 2\lambda_2}\right),\label{srwsringsur}\\
\mathcal L&\simeq&1+\frac{2\epsilon}{\pi}\,\lambda_2\tanh\lambda_2\,,\label{lrsur}\\
1-\nu&\simeq&\frac{2\epsilon^3}{\pi}\,\tanh\lambda_2\left(\frac{\lambda_2}{\cosh^2\lambda_2}+\tanh\lambda_2\right)\!.\label{nursur}
\end{eqnarray}
Equations (\ref{lrsur}) and (\ref{nursur}) show that $\lambda_2$ diverges in this limit as $\lambda_2\sim \left(\mathcal L-1\right)/\left(1-\nu\right)^{1/3}$. Inserting this asymptotic into (\ref{qrwsring}) we see that the optimal density profile decays exponentially outside of the interval $|x|<1$ over a length scale proportional to $\left(1-\nu\right)^{1/3}$. For not too small $\mathcal L$, this decay length is much smaller than the length of the complementary interval, $\left(1-\nu\right)^{1/3}\ll\mathcal L -1$. Thus when $1-\nu\ll\left(\mathcal L-1\right)^3$, the optimal profile is exponentially localized, and we can expect that the sub-leading expression for the action in this limit is independent of the ring size. Indeed, using the above expression for $\lambda_2$ and Eq.~(\ref{nursur}) in Eq.~(\ref{srwsringsur}) we obtain the asymptotic (\ref{highring}) which coincides with the corresponding asymptotic for the infinite line. Using the duality relation~(\ref{dual2}), we obtain the void asymptotic (\ref{highring2}) as well.

\subsection{Three asymptotics}

Finally we present three asymptotics of the occupation statistics in terms of the function $g(\nu,\mathcal L)$.
Close to the average occupation fraction, $|\nu-\bar{\nu}|\ll \bar{\nu}$, we get
\begin{equation}
\label{lowring}
g(\nu,\mathcal L)\simeq \frac{3}{2}\frac{\left(\nu-\bar{\nu}\right)^2\mathcal L^2}{\left(\mathcal L-1\right)^2}\,.
\end{equation}
When $1-\nu\ll \text{min}\left[1,    \left(\mathcal L-1\right)^3 \right]$, we obtain
\begin{equation}
\label{highring}
\frac{\pi^2}{2} - g(\nu,\mathcal L)\simeq \frac{3\pi^{4/3}}{2^{1/3}}\left(1-\nu\right)^{1/3}.
\end{equation}
Finally in the $\nu\ll \text{min}\left[1,\left(\mathcal L-1\right)^{-3} \right]$ range
\begin{equation}
\label{highring2}
g(\nu,\mathcal L)\simeq \frac{1}{\left(\mathcal L-1\right)^2}\left(\frac{\pi^2}{2}-\frac{3\pi^{4/3}}{2^{1/3}}\nu^{1/3}\right).
\end{equation}
The variance is obtained by using Eqs.~(\ref{lowring}) and~(\ref{ringactionrws}) to yield, in agreement with \cite{berez},
\begin{equation}\label{var}
\text{Var}\left(\nu \right)=\frac{2l^2\left(L-l\right)^2}{3NL^2D_0T},
\end{equation}

\


\begin{thebibliography} {99}

\bibitem{levi}
P. L\'{e}vy, Comp. Math. \textbf{7}, 238 (1939).
	
	
\bibitem{Levy}
	P.~L\'{e}vy,
	{\it Processus Stochastiques et Mouvement Brownien}
	(Gauthier-Villars, Paris, 1948).
	
\bibitem{oldsurv}
M. Kac, Proc. Second
Berkeley Sympos. on Math. Statist. and Prob. (Univ. California Press, Berkeley, 1951).
	
\bibitem{kac}
D. A. Darling and M. Kac, Trans. Amer. Math. Soc. \textbf{84},  444 (1957).	

\bibitem{Ray}
D. Ray, Illinois J. Math. \textbf{7}, 615 (1963).


\bibitem{Knight}
F. B. Knight, Trans. Am. Math. Soc. \textbf{109}, 56 (1963).


\bibitem{spi}
F. Spitzer, {\it Principles of Random Walk} (Van Nostrand, Princeton, N.J., 1964).

\bibitem{IM65}
K.~It\^{o} and H.~P.~McKean,
{\it Diffusion Processes and Their Sample Paths}
(New York, Springer, 1965).

\bibitem{noam1}
N. Agmon, J. Chem. Phys. \textbf{81}, 3644 (1984).
	
\bibitem{noam2}
	A. M. Berezhkovskii, V. Zaloj and N. Agmon,
	Phys. Rev. E \textbf{57}, 3937 (1998).
	
\bibitem{satya}
S. N. Majumdar and A. Comtet, Phys. Rev. Lett. \textbf{89}, 060601 (2002).

\bibitem{many}
O. B\'{e}nichou, M. Coppey, J. Klafter, M. Moreau and G. Oshanin, J. Phys. A \textbf{36},  7225 (2003).

\bibitem{simtwo}
O. B\'{e}nichou, M. Coppey, J. Klafter, M. Moreau and G. Oshanin, J. Phys. A \textbf{38},  7205 (2005).

\bibitem{reactions}
O. B\'{e}nichou, M. Coppey and M. Moreau, J. Chem. Phys. \textbf{123}, 194506 (2005).



\bibitem{eli}
E. Barkai, J. Stat. Phys. \textbf{123}, 883 (2006).

\bibitem{greben}
D. S. Grebenkov, Phys. Rev. E. \textbf{76}, 041139 (2007).

\bibitem{benyopty}
O. B\'{e}nichou and R. Voituriez,
J. Chem. Phys. \textbf{131}, 181104 (2009).

\bibitem{benyopty2}
 O. B\'{e}nichou and J. Desbois, J. Phys. A \textbf{42}, 015004 (2009).

\bibitem{BM:book}
P. M\"{o}rters and Y. Peres, {\it Brownian Motion}
(Cambridge University Press, Cambridge, 2010).

\bibitem{berez}
A. M. Berezhkovskii, Chem. Phys. \textbf{370}, 253 (2010).

\bibitem{noam3}
N. Agmon, Chem. Phys. Lett.  \textbf{497}, 184 (2010).

 \bibitem{beni}
 O. B\'{e}nichou and R. Voituriez,
Phys. Rep. \textbf{539}, 225 (2014).

\bibitem{singlefile}
O. B\'{e}nichou and J. Desbois, J. Stat. Mech. (2015) P03001.

\bibitem{tridib1}
 P. L. Krapivsky, K. Mallick and T. Sadhu, J. Stat. Mech.  (2015) P09007.

\bibitem{Hugo16}	
F. Angeletti  and H. Touchette, J. Math. Phys. \textbf{53}, 023303	(2016).

\bibitem{Hugo16b}
P. Tsobgni Nyawo and H. Touchette, Europhys. Lett. {\bf 116}, 50009 (2016).

\bibitem{Hugo18}
P. Tsobgni Nyawo and H. Touchette, Phys. Rev. E \textbf{98}, 052103 (2018).

\bibitem{sid}
J. Randon-Furling and S. Redner, J. Stat. Mech. (2018) P103205.


\bibitem{bress}
P. C. Bressloff and J. M. Newby, Rev. Mod. Phys. \textbf{85}, 135 (2013).

\bibitem{MVK}
B. Meerson, A. Vilenkin and P. L. Krapivsky, Phys. Rev. E \textbf{90}, 022120 (2014).


\bibitem{surv}
T. Agranov, B. Meerson and A. Vilenkin,
Phys. Rev. E \textbf{93}, 012136 (2016).

\bibitem{AMabsorption}  T. Agranov and B. Meerson, Phys.  Rev.  E \textbf{95}, 062124 (2017).


\bibitem{narrow}
T. Agranov and B. Meerson,
Phys. Rev. Lett. \textbf{120}, 120601 (2018).

\bibitem{AMreview} T.  Agranov  and  B. Meerson, in \textit{``Chemical Kinetics
Beyond the Textbook"}, edited by K. Lindenberg, R. Metzler and G. Oshanin (World Scientific, Singapore, 2018), chapter 9; arXiv:1806.02114.	

\bibitem{MFTreview} L. Bertini, A. De Sole, D. Gabrielli, G. Jona Lasinio, C. Landim. Rev. Mod. Phys. \textbf{87}, 593 (2015).

\bibitem{ber}
L. Bertini, A. De Sole, D. Gabrielli, J. Jona-Lasinio and C. Landim, J. Stat. Phys. \textbf{123}, 237 (2006).

\bibitem{main2}
T. Bodineau and B. Derrida, C. R. Physique \textbf{8}, 540 (2007).

\bibitem{hurtado}
P. I. Hurtado, C. P. Espigares,
J. J. del Pozo, and P. L. Garrido, 	J. Stat. Phys. \textbf{154}, 214 (2014).


\bibitem{main}
O. Shpielberg and E. Akkermans, Phys. Rev. Lett. \textbf{116}, 240603 (2016).

\bibitem{phase}
Y. Baek, Y. Kafri and V. Lecomte,
Phys. Rev. Lett. \textbf{118}, 030604 (2017).

\bibitem{hugo2009}
H. Touchette, Phys. Rep. \textbf{478}, 1 (2009).

	
\bibitem{Paulbook}
P. L. Krapivsky, S. Redner and E. Ben-Naim, \textit{A Kinetic View of Statistical Physics}
(Cambridge University Press, Cambridge, 2010).

\bibitem{notehugo}
Refs. \cite{Hugo18,Hugo16b} also accounted for a constant driving force and discovered a dynamical phase transition resulting from it. See also earlier work \cite{Nieuwenhuizen,Grassberger}, where a constant driving force causes a dynamical phase transition in a different setting.

\bibitem{DonskerVaradhan}
M. D. Donsker and S. R. S. Varadhan,
Comm. Pure Appl. Math.  \textbf{28}, 1 (1975); \textbf{28}, 279 (1975);
\textbf{29}, 389 (1976);
\textbf{36}, 183 (1983).


\bibitem{level2}
J. Hoppenau, D. Nickelsen and A. Engel, New J. Phys. \textbf{18}, 083010 (2016).


\bibitem{conditioned}
R. Chetrite and H. Touchette, Phys. Rev. Lett. \textbf{111}, 120601 (2013).

\bibitem{conditioned2}
R. Chetrite and H. Touchette, J. Stat. Mech. (2015) P12001.

\bibitem{Spohn}
H. Spohn, \textit{Large-Scale Dynamics of Interacting Particles} (Springer-Verlag, New York, 1991).
	

	
\bibitem{KL}
C. Kipnis and C. Landim, \textit{Scaling Limits of Interacting Particle Systems} (Springer, New York, 1999).



\bibitem{meso}		
Ya. M. Blanter and M. B\"{u}ttiker, Phys. Rep. \textbf{336}, 1 (2000).
		
		

		
\bibitem{jor}		
A. N. Jordan, E. V. Sukhorukov,  and S. Pilgram, J. Math. Phys. \textbf{45}, 4386 (2004).


\bibitem{penini}
R. Pnini and B. Shapiro, Phys. Rev. B \textbf{39}, 6986 (1989).

\bibitem{shapiro2}
R. Sarma, A. Yamilov, P. Neupane, B. Shapiro, and H. Cao,
Phys. Rev. B \textbf{90}, 014203 (2014).


\bibitem{Liggett}
T. M. Liggett, \textit{Stochastic Interacting Systems: Contact, Voter,
	and Exclusion Processes} (Springer, New York, 1999).


\bibitem{Chou99}
T. Chou and D. Lohse, Phys. Rev. Lett. \textbf{82}, 3552 (1999).

\bibitem{Das10}
A. Das, S. Jayanthi, H.S.M.V. Deepak, K. V. Ramanathan, A. Kumar, C. Dasgupta and A. K. Sood, ACS Nano {\bf 4}, 1687 (2010).

\bibitem{oshanintracer}
O. B\'{e}nichou, P. Illien, G. Oshanin, A. Sarracino and R. Voituriez, J. Phys.: Cond. Matter \textbf{30}, 443001 (2018).

\bibitem{cell}
T. Chou, K. Mallick, and R. K. P. Zia, Rep. Prog. Phys. \textbf{74}, 116601 (2011).

\bibitem{zrpbook}
A. Schadschneider, D. Chowdhury and K. Nishinari \textit{Stochastic Transport in Complex Systems} (Elsevier Science, 2010).

\bibitem{void}

P. L. Krapivsky, B. Meerson and P. V. Sasorov, J. Stat. Mech. P12014 (2012).


\bibitem{sad}
In the occupation statistics problem one should demand that the number of particles,  which occupy the macroscopic interval of length $2l$, be large: $N\nu\gg1$.


\bibitem{bd}

T. Bodineau and B. Derrida, Phys. Rev. Lett. \textbf{92}, 180601 (2004).

\bibitem{Itwould}
It would be interesting to find a lattice gas model, for which this condition is violated. If this happens, the total number of particles, composing the stationary optimal density profile, should come from an additional minimization procedure.

\bibitem{Convergence}
Convergence of the integral (\ref{transform1})
puts some limitations on the behavior of $D(\rho)$ and $\sigma(\rho)$ at small densities. As an example, let $D(\rho\rightarrow0)\sim\rho^{\alpha}$ and $\sigma(\rho\rightarrow0)\sim\rho^{\beta}$. Then the integral converges at $\rho\rightarrow0$ if and only if $2\alpha-\beta+2>0$. This condition holds for the models we consider here.


\bibitem{tangent}
 L. Elsgolts, \textit{Differential Equations and the Calculus of Variations} (Mir Publishers, Moscow, 1977).


\bibitem{quant}
S. Fl\"{u}gge, \textit{Practical Quantum Mechanics} \RomanNumeralCaps{1} (Springer-Verlag, Berlin, 1971).

\bibitem{landau}
L.D. Landau and E.M. Lifshits, \textit{Quantum Mechanics, Non-Relativistic theory} (Pergamon, London, 1958).

\bibitem{For}
For solutions with compact support it vanishes for all points past the support ends $|x|>x_0$.

\bibitem{Actually}  Equations~(\ref{conti}) and (\ref{rwsconst}) have infinitely many solutions, where $\lambda_1$ is on the intervals $\pi m\leq\lambda_1\leq\pi m+\pi/2$;  $m=0,1,2,\dots$.
The values $m=1,2,\dots$, correspond to the excited states of the quantum problem and are suboptimal.






\bibitem{Wellipticfunctions} Wolfram Research, Inc.,\\ http://functions.wolfram.com/EllipticFunctions/.

\bibitem{Wellipticintegrals} Wolfram Research, Inc.,\\ http://functions.wolfram.com/EllipticIntegrals/.

\bibitem{Reintroducing}
Reintroducing the lattice constant $a$, we obtain a dimensionally correct formula
$-\ln\mathcal P(\nu,N,T)\simeq [D_0T/(\ell a)]\,g\left(na,\nu\right)$.

\bibitem{fornu}
For $\nu>\nu_{c1}(\mathcal L)$ the solution (\ref{u1zr}) would violate the positivity constraint (\ref{positive}) and is therefore invalid, so the solution (\ref{123}) is unique there.


 \bibitem{land}
 H. E. Stanley, \textit{Introduction to Phase Transitions and Critical
 Phenomena} (Oxford University Press, Oxford, 1971).

\bibitem{Nieuwenhuizen}
Th. M. Nieuwenhuizen, Phys. Rev. Lett. \textbf{62}, 357 (1989).

\bibitem{Grassberger}
V. Mehra and P. Grassberger, Physica  D  \textbf{168–-169}, 244 (2002).

\bibitem{flor}
 J.R. Lakowicz,
 \textit{Topics in Fluorescence Spectroscopy} (Springer, Boston, MA, 2002).

 \bibitem{KrMe} P. L. Krapivsky and B. Meerson,  Phys. Rev. E \textbf{86}, 031106 (2012).

\end{thebibliography}
\end{document}